\documentclass[10pt,preprint2]{aastex}

\usepackage{amsmath}

\newcommand{\mydeg}{{$^{\circ}$}}

\newcommand{\etal}{{\it et~al.}}

\begin{document}

\title{Surveying the Inner Solar System with an Infrared Space Telescope}

\shorttitle{NEA Survey}
\shortauthors{Buie, Reitsema, \& Linfield}

\author{Marc W. Buie}
\affil{Southwest Research Institute, 1050 Walnut St., Suite 300, Boulder, CO~~80302}
\affil{B612 Foundation, 20 Sunnyside Ave., Suite 427, Mill Valley, CA~~94941}
\email{buie@boulder.swri.edu}

\author{Harold J. Reitsema}
\affil{B612 Foundation, 20 Sunnyside Ave., Suite 427, Mill Valley, CA~~94941}
\email{harold@b612foundation.org}

\author{Roger P. Linfield}
\affil{B612 Foundation, 20 Sunnyside Ave., Suite 427, Mill Valley, CA~~94941}
\email{rplinfield@comcast.net}

\slugcomment{Version 1.31, submitted to AJ 2016/02/02, accepted 2016/07/11}

%\pagebreak
%\clearpage
%\cleardoublepage

\begin{abstract}

We present an analysis of surveying the inner Solar System for objects
that may pose some threat to the Earth.  Most of the analysis is based
on understanding the capability provided by Sentinel, a concept for an
infrared space-based telescope placed in a heliocentric orbit near the
distance of Venus.  From this analysis, we show 1) the size range being
targeted can affect the survey design, 2) the orbit distribution of the
target sample can affect the survey design, 3) minimum observational arc
length during the survey is an important metric of survey performance,
and 4) surveys must consider objects as small as $D=15-30$~m to meet
the goal of identifying objects that have the potential to cause damage
on Earth in the next 100 years.  Sentinel will be able to find 50\% of
all impactors larger than 40 meters in a 6.5 year survey.  The Sentinel
mission concept is shown to be as effective as any survey in finding
objects bigger than $D=140$~m but is more effective when applied to
finding smaller objects on Earth-impacting orbits.  Sentinel is also
more effective at finding objects of interest for human exploration that
benefit from lower propulsion requirements.  To explore the interaction
between space and ground search programs, we also study a case where
Sentinel is combined with the Large Synoptic Survey Telescope and show the
benefit of placing a space-based observatory in an orbit that reduces the
overlap in search regions with a ground-based telescope.  In this case,
Sentinel$+$LSST can find more than 70\% of the impactors larger than 40
meters assuming a 6.5 year lifetime for Sentinel and 10 years for LSST.

\end{abstract}

\keywords{}

\section{Introduction}

Near Earth Asteroids (NEAs) are a population of asteroids that spend
at least part of the time in the inner solar system that are both
potential targets for future exploration missions and possible threats
of Earth impact.  Surveys for NEAs, notably NEAT \citep{pra99}, LINEAR
\citep{sto00}, Pan-STARRS \citep{jed03,jed06} and the Catalina Sky
Survey \citep{lar07}, have found over 90\% of NEAs larger than about 1
km in diameter and a total of over 13,000 of all sizes \citep{jed15}.
\citet{ted00} and \citet{cel04} presented the idea of of space-based
infrared survey instruments and the Earth-orbiting WISE telescope
has now demonstrated NEA detection in infrared wavelengths from
space, adding about 200 additional NEA discoveries \citep{mai15b}.
Because of the power-law distribution of NEA sizes and following a
recent re-examination of the population of small NEAs \citep{bos15},
current estimates show that there are many times as many small NEAs
that have not been found, perhaps as many as 4 million larger than 30
meters that could cause substantial damage upon impact. \citet{che04}
and \citet{ver09} considered the probability of impact of these small
objects and demonstrated that impactors are likely to come from a limited
range of orbital parameters, providing a sub-population of NEAs of most
interest to planetary defense.  This interest motivates the drive for
better NEA searches for planetary defense. The National Research Council
in 2010 recommended that ``surveys should attempt to detect as many 30-
to 50-meter objects as possible'' \citep{nrc10}.  Such searches will also
provide numerous targets that are within reach of available exploration
mission launch and rendezvous capabilities.

Current surveys, while making clever and effective use of their
facilities, are adding only about 2000 new discoveries per year.  The rate
of detection will rise by two orders of magnitude with the 2022 completion
of the Large Synoptic Survey Telescope \citep[LSST,][]{ive07,ive14}, which
claims the potential to reach 82\% completion on PHAs larger than 140
meters in 10 years.  The LSST survey completion rate for smaller objects
will be lower, 34\% completion in 10 years down to Tunguska-like
40-meter objects.  The Sentinel mission \citep{lu13,rei15} is proposed
to search for NEAs from a heliocentric orbit near Venus' distance from
the Sun (0.7 Astronomical Units, AU).  As will be shown in this work,
Sentinel surveys a unique volume of space and produces a complementary
set of objects with a discovery rate even higher than that of LSST.

The goal of the present work was to understand how an infrared
space-based observatory can best be used to extend present-day surveys
to a smaller size range for discovery of future impact threats.
Working in the thermal infrared provides advantages for detecting NEAs
since the target flux depends mostly on its heliocentric distance and
much less on its albedo than does the visible flux.  This property of an
infrared survey provides a complementary aspect when run in parallel with
optical surveys.  Because of this weak dependence on albedo, infrared
flux can more readily be used to deduce the size of the objects.  Also,
the phase angle dependence of brightness is less for thermal infrared
observations than for visible reflectance.  Additionally, the background
source density is lower in the infrared, reducing difficulties due to
confusing objects with background sources.

Presented here is an analysis of a series of mission options that
include the baseline concept for the Sentinel Mission \citep{lu13,rei15}.
There is no attempt here to demonstrate an optimal mission architecture:
there are likely to be other designs that can work as well.  The purpose
of this work was to investigate Sentinel and see if it is sufficient
to the task and also probe potentially beneficial variations in the
mission design.  Our statistical modeling method allows very fast
execution and permits testing many different scenarios without requiring
true exposure-by-exposure fidelity of a simulated detection dataset.
Using this tool, the Sentinel design will be shown to be a very good
approach for finding hazardous objects.

\section{Baseline Mission Profile}

The core of this work is an analysis of one particular mission design,
called the baseline mission profile which represents the nominal
properties of the Sentinel observatory.  The top-level characteristics
of the baseline mission are summarized in Table~\ref{tbl-specs}.
The observatory is a wide-field infrared, fully steerable space-based
telescope.  The detectors employ Mercury-Cadmium-Telluride (HgCdTe)
photo-sensitive material bonded to Capacitive Trans-Impedance Amplifier
(CTIA) readout circuits.

\begin{deluxetable}{cccccccccc}
\tablecaption{Sentinel -- Baseline Capabilities\label{tbl-specs}}
\tablewidth{0pt}
\tablehead{
\colhead{Item} &
\colhead{Value} &
\colhead{Notes}
}
\startdata
Telescope Aperture& 50 cm unobstructed& Passively cooled to 60K\cr
Pixel Scale&  2.15 arcsec/pix& Critically sampled on NEAs\cr
Read noise& 110 e-/pixel& per readout\cr
Dark current& 600 e-/sec& Active closed-cycle cooling to 30K\cr
Field of View& 2\mydeg\ by 5.5\mydeg& 2x5 detector mosaic\cr
Fill Factor& 96\%& small gaps between detectors\cr
Wavelength& 5 -- 10.2 $\mu$m& Unfiltered, set by detector\cr
Exposure& 180 sec& six 30-sec images\cr
Field of Regard& Solar Elongation $>$ 80\mydeg& set by sunshade\cr
Observing Cycle& 28 days& covers FOR four times\cr
Cycle Cadence& 0, 1h, 48h, 49h& two pairs\cr
Semi-major axis& 0.66 AU& Similar to Venus\cr
Eccentricity& 0.091& Not critical\cr
Inclination& 0.27\mydeg& Not critical\cr
\enddata
%\tablecomments{
%The simulation and conversion from $D$ to $H_V$ was based on a fixed albedo
%of 14\%.
%}
\end{deluxetable}

This choice of orbit was a consequence of optimizing the ability
to detect NEAs.  The orbit is also practical to achieve by taking
advantage of a Venus gravity assist during the early phase of the
mission.  The orbital elements we use in modeling the baseline (V for
Venus-like orbit) are nominal values used by the B612 Foundation's
industrial partner, Ball Aerospace, in their design work.  The gravity
assist is used to lower the aphelion distance and period of the orbit
and the values shown in Table~\ref{tbl-specs} reflect nominal values.
The observatory is designed to function at any distance between Venus and
Earth and the final orbit need not match the design values precisely.
The angular elements used for the simulation are arbitrary.  Also, we
make no attempt to model the portion of the mission that will execute
prior to the Venus flyby although routine survey operations are planned
during the transit to Venus.  The survey calculations assume the notional
orbit for the entire duration of the mission.

The field of regard (FOR) covers slightly more than half the sky.
The imaging system can cover this region 4 times in a 28-day observing
cycle.  The chosen observing location does come at a price.  The distance
from Sentinel to the Earth is too large for transmitting all data back
for analysis, similar to the design of the Kepler Mission \citep{koc10}.
Thus this design requires significant on-board processing to identify
the moving objects in the data and then specify small regions of interest
(ROI) around the moving objects that will be transmitted to the ground.
In reality, the on-board data processing is limited to the detection
of transients in the image data that by its design will consist largely
of moving objects.  This on-board processing significantly reduces the
volume of data that must be transmitted.  These ROIs are the primary
dataset from which final astrometry, linkages, and orbit estimations
are generated in the ground system.  More details about this aspect of
the mission design can be found in \citet{lu13,rei15}.

In our work, we describe the observations of an object as a 4-observation
set collected in a 28-day observing cycle.  This is an accurate
description but can be mis-interpreted.  Each observation is the result of
combining six independent, back-to-back, images.  These six ``sub-images''
are combined with a robust mean estimator that will eliminate any
non-statistical outlier pixels.  Only then is the combined image scanned
for a detection above our signal-to-noise ratio threshold.  Thus,
these basic set of observations is really a 24-observation set where
data have been aggregated down by a factor of six before transmission to
the ground.  While we discuss these observations as a pair of duples,
there is considerably more data at work here.  This method can deliver
high-confidence detections with very low false positive rates provided
that the detectors are sufficiently stable.  For the purposes of this
study, we assume stability but this is clearly an issue that we must
track very carefully during development of flight hardware.

\section{Delivered Orbit Quality}

An important outcome of the survey is not just a list of detected
objects but linked observations of objects across multiple observing
epochs and an estimate of their orbits.  In the survey simulator, the
number of observations of an object is tracked by counting the number
of successful observing cycles and the time-span between the first
and last cycle.  A successful observing cycle is defined as one where
the object is detected 4 times out of 4 opportunities within a cycle.
If the object is detectable in an observing cycle but fails to be counted,
the dominant reason for a failure is having the object fall on a chip gap
in one or more observations.  Generally, the variation in detectability
from flux or target motion is small within a single observing cycle.

An essential component of a survey is clearly the ability to link
observations from one detection to another.  It was beyond the scope of
this work to address the complete linkage problem.  Instead, we make a
few simplifying assumptions.  The most important of these assumptions
is that observations between different observing cycles can be linked.
With the Sentinel cadence design, one cycle yields a 4-measurement track
with a 2-day arc.  Armed with the linkage assumption, we computed test
cases that illustrate the orbit estimation quality as a function of the
available data on an object.

For the following computations, a few sample object orbits were
constructed.  Given a set of observing times and an orbit, the spacecraft
and object position are computed.  To this is added measurement noise
of 1.3 arc-sec per observation.  This noise level is equal to 0.5
pixels and is estimated to be a conservative uncertainty that includes
all sources of error from such components as centroiding, astrometric
image solutions, and spacecraft location.  These noise-injected values
constitute a synthetic astrometric dataset that can be used to fit
an orbit.  Our tools are based on OpenOrb \citep{gra09}.  For short-arc
fits we use the OpenOrb statistical ranging Monte Carlo approach for the
orbit and error estimations.  Once the observation arc is long enough
we can generate orbit fits and error estimation with the least-squares
Gaussian estimation in OpenOrb.  In this work we chose to use the Monte
Carlo approach for data spanning 60 days or less and least-squares for
longer arcs.  We found that 60 days was the longest arc where statistical
ranging was useful and used that even though the least-squares fitting
also worked equally well over the high end of that range.

\subsection{Short-arc data ($\le$1 hour arc)}

The first case to consider is the accuracy of the orbit estimate and the
ability to predict future positions given the first pair of observations.
For the baseline Sentinel design, these two points will be separated by
one hour.  Table~\ref{tbl-short} contains a summary of the predictive
power of this short-arc observation.  The uncertainties are shown after
converting to standard (tangent plane) coordinates, using the best-fit
position as the tangent plane point -- $\xi$ is in the direction of
right ascension (or ecliptic longitude) and $\eta$ is in the direction
of declination (or ecliptic latitude).  Most of the Monte Carlo orbits
generate positions in the same general area.  However, some of the objects
evolve significantly away from the cloud.  These objects are those that
happen to be very near the observatory at the time of observation and
their future position can change very quickly.  The implication from
Table~\ref{tbl-short} is that an object needs to be seen again within
a few days or it will be quickly lost.  Note that the uncertainty in
the position at the next planned observation in the survey cadence is
roughly 4~arcmin (2 day predict).  Based on our prior experience, the
linkage to a second should be straightforward but simulations like these
can be very useful to fine-tune the observing cadence if difficulties
in making solid linkages arise.

\begin{deluxetable}{cccl}
\tablecaption{Short-arc (1-hour) data -- Future Predictions\label{tbl-short}}
\tablewidth{0pt}
\tablehead{
\colhead{Predict} &
\colhead{$\sigma_{\xi}$} &
\colhead{$\sigma_{\eta}$} &
\colhead{Notes}\cr
& (arcsec)& (arcsec)&
}
\startdata
 1 sec& 2.8& 2.8& Region looks square, no outliers\cr
 1 min& 2.8& 2.8& Same region as for 1 sec predict\cr
 5 min& 3.0& 3.0& Region still looks square\cr
10 min& 3.3& 3.3& Region still square, isolated outliers\cr
15 min& 3.5& 3.6& Worst outliers at 15 arcsec\cr
30 min& 4.4& 4.4& Region starting to round, worst at 30 arcsec\cr
  1 hr& 6.8& 6.2& Region even rounder, worst at 100 arcsec\cr
  2 hr&  10&  10& Region circular, worst at 800 arcsec\cr
  4 hr&  18&  18& worst at 34 arcmin\cr
 1 day& 104& 110& \cr
 2 day& 229& 251& Time of next observation\cr
 4 day& 572& 618& \cr
\enddata
\tablecomments{\scriptsize
This table shows one standard deviation of the sky-plane position
($\sigma_{\xi},\sigma_{\eta}$)
at different times (Predict) from the last observation.
The notes describe the shape or other attributes of the uncertainty
region of the object based on the orbit fit.  Astrometric uncertainty
assumed to be 1.3 arcsec per observation.
}
\end{deluxetable}

\subsection{Single Observing Cycle (2 day arc)}

The second case is the orbit estimation and future position predictions
based on a 4-observation set from a single observing cycle.  In this case,
the time spread in the observations is 2 days.  Table~\ref{tbl-twoday}
lists the predicted uncertainty versus the time since the last
observation.  The uncertainty region at short prediction times is
a circular region dominated by the size of the measurement errors.
At longer prediction times the shape of the region is elliptical with a
considerably smaller number of outlier objects than was the case for the
1-hour arc data.  At the time of the next observing cycle, the positional
uncertainty is substantial and linkages between the two observing cycles
will need to use more than a positional coincidence between the object
and the prediction to support the linkage.  Additional information,
such as orbit pole coincidence, rate of motion, direction of motion,
as well as position will be required to make the linkage.

\begin{deluxetable}{cccl}
\tablecaption{2-day arc data -- Future Predictions\label{tbl-twoday}}
\tablewidth{0pt}
\tablehead{
\colhead{Predict} &
\colhead{$\sigma_{\xi}$} &
\colhead{$\sigma_{\eta}$} &
\colhead{Notes}\cr
& (arcsec)& (arcsec)&
}
\startdata
 1 sec& 2.1& 2.0& circular uncertainty region\cr
 1 min& 2.1& 2.0& Same region as for 1 sec predict\cr
  1 hr& 3.0& 4.1& uncertainty region slightly non-circular\cr
  8 hr&  14&  18& uncertainty region definitely elongated\cr
 1 day&  51&  88& \cr
 8 day& 1279& 1897& \cr
26 day& 2.4\mydeg& 3.2\mydeg& Time of next observation\cr
\enddata
\tablecomments{\scriptsize
This table shows the sky-plane uncertainty ($\sigma_{\xi},\sigma_{\eta}$)
at different times (Predict) from the last observation.
The notes describe the shape or other attributes of the uncertainty
region of the object based on the orbit fit.  Astrometric uncertainty
assumed to be 1.3 arcsec per observation.
}
\end{deluxetable}

\subsection{Multi-Cycle arc ($\ge$28 days)}

If an object is seen in two adjacent observing cycles, the orbit estimate
gets much better.  In OpenOrb, the orbit fitting can either be done
with statistical ranging or traditional least-squares Gauss solutions.
Table~\ref{tbl-twocycle} shows how the positional uncertainty holds
up over time for two observations in two adjacent observing cycles
with a 28-day arc.  At this point, the orbit estimate is becoming
quite good and linkages to future observations is straightforward.
Once three consecutive cycles are linked the 10-year prediction is good
to $\sigma_{\xi}=37''$, $\sigma_{\eta}=12''$.  Thus the NEA position from
3 observing cycles is sufficient to permit targeted followup observations
for physical characterization and high-precision orbit estimates.

\begin{deluxetable}{cccl}
\tablecaption{28-day (2-cycle) arc data -- Future Predictions\label{tbl-twocycle}}
\tablewidth{0pt}
\tablehead{
\colhead{Predict} &
\colhead{$\sigma_{\xi}$} &
\colhead{$\sigma_{\eta}$} &
\colhead{Notes}\cr
& (arcsec)& (arcsec)&
}
\startdata
  1 hr& 1.5& 1.7& Error cloud dominated by measurement error\cr
 1 day& 2.0& 2.6& Worst error is 7 arcsec\cr
 2 day& 2.5& 3.7& Error clound becoming elliptical\cr
28 day&  30&  69& Elliptical error region\cr
56 day&  92& 183& error region similar to a line-of-variation\cr
  1 yr& 636&  13& \cr
\enddata
\tablecomments{\scriptsize
This table shows the sky-plane uncertainty ($\sigma_{\xi},\sigma_{\eta}$)
at different times (Predict) from the last observation.
The notes describe the shape or other attributes of the uncertainty
region of the object based on the orbit fit.  Astrometric uncertainty
assumed to be 1.3 arcsec per observation.
}
\end{deluxetable}

\subsection{Alternate Strategies}

Table~\ref{tbl-basecad} summarizes the predicted positional uncertainty
at the next epoch based on the data in hand for the baseline cadence.
The column labeled N$_{\rm obs}$ provides the number of individual
observations of the object.  N$_{\rm cycles}$ indicates the number of
observing cycles needed to get these observations.  The column labeled
``arc'' gives the shortest possible arc-length of the observations.
Note that spreading out a multi-cycle dataset for a longer arc with the
same number of cycles will provide a better orbit estimate.  The column
labeled ``predict'' gives the length of time before the next possible
observing opportunity.  The final two columns give the uncertainty in the
position for the time of the predict interval from the last observation.
A special case is shown in the last line of a prediction for three-cycle
dataset at 10 years from the observations.

\begin{deluxetable}{cccccc}
\tablecaption{Baseline Cadence Linkage Summary\label{tbl-basecad}}
\tablewidth{0pt}
\tablehead{
\colhead{N$_{\rm obs}$} &
\colhead{N$_{\rm cycles}$} &
\colhead{arc} &
\colhead{predict} &
\colhead{$\sigma_{\xi}$} &
\colhead{$\sigma_{\eta}$}\cr
& & & & (arcsec)& (arcsec)
}
\startdata
  2& 0.5&   1 hr&  2 day& 229& 251\cr
  4&   1&  2 day& 26 day& 8640& 11520\cr
  8&   2& 28 day& 28 day&  30&  69\cr
 12&   3& 56 day& 28 day& 2.0& 3.7\cr
 12&   3& 56 day& 10 year& 37&  12\cr
\enddata
\tablecomments{\scriptsize
This table shows the sky-plane uncertainty ($\sigma_{\xi},\sigma_{\eta}$)
at the time each linkage is made in extending the arc from discovery into
being catalog.  N$_{\rm obs}$ is the cumulative number of observations over
N$_{\rm cycles}$, where one cycle is 28-days that are used support each
linkage.  ``Arc'' gives the length of the constraining arc and ``predict''
gives the extrapolation time from last linked observation to the new data
being linked.  Astrometric uncertainty was assumed to be 1.3 arcsec per
observation.
}
\end{deluxetable}

The results in Table~\ref{tbl-basecad} can support an investigation
into alternate cadences.  The most important element of the survey
process illuminated by this part of the analysis is the challenge of
accurate linkages.  The first linkage within the first observing cycle
can be made easier by reducing the time between the first two pairs.
If this time is reduced to 1 day, the positional uncertainty is reduced
to ($\sigma_{\xi},\sigma_{\eta})=(127'',155'')$ but this more than
doubles the uncertainty at the next observing cycle, assuming the total
number of observations remains constant.  The only way to reduce the
uncertainty at the time of the first cross-cycle linkage is to obtain more
observations within an observing cycle.  Making such a change affects
the overall cadence but for this computation these changes are ignored.
If an observing cycle were built from three pairs each separated by 30
minutes, the uncertainty at the next cycle is reduced by a factor of two.
The cost of such a change is considerable and must be traded off against
the need to make the linkage process easier.  Discussions with the Minor
Planet Center indicated that linkages even with the baseline cadence
would be routinely possible and we did not pursue this issue any further.
This assumption may well be worth further study but is beyond the scope
of this work.

\section{Detection}

The survey simulator calculates the amount of collected thermal radiation
from the NEA\null.  If this is high enough above the background noise
then the object can be detected.  The signal collected is a combination of
the thermal radiation and smearing from the PSF and any trailing losses.
We ignore any contribution to reflected solar light in our passband.

\subsection{Thermal Emission Model}

The model for the thermal emission detected is a simple approximation to
the problem.  In general, the NEAs have a distribution of spin rates,
rotation poles, and surface regolith properties but these details
are ignored in favor of reasonable average values.  The first step
is to compute the isothermal surface temperature for the NEA based on
radiative equilibrium.  We used the thermal balance approximation
for the temperature \citep[eg.][]{spe90} given by
\begin{equation}\label{eq-tnea}
T_{\rm NEA} = \left( {(1-A) L_\sun \over r^2 4 \epsilon \sigma } \right)
^ {1 \over 4}
\end{equation}
where $A$ is the bond albedo, $L_\sun$ is the solar constant
(1.374 $\times$ 10$^{-5}$ erg cm$^{-2}$ s$^{-1}$ from \citet{wil80}),
$r$ is the heliocentric distance [AU], $\epsilon$ is the emissivity of
the surface, and $\sigma$ is the Stefan-Boltzmann constant
[erg cm$^{-2}$ s$^{-1}$ K$^{-4}$].  The numeric factors in the denominator
are based on the conservative choice of an isothermal model.
The bond albedo is estimated from the geometric albedo using
the phase integral from \citet{bow89} and a phase coefficient of $G=0.2$.
The model
approximates the variation in surface temperature from the sunlit
and dark sides by the simple expression
\begin{equation}\label{eq-tbb}
T_{\rm BB} = T_{\rm NEA} + \Delta T \cos(\alpha)
\end{equation}
where $\Delta T$ is a temperature correction factor and $\alpha$ is the
solar phase angle.  This expression gives the effective black-body temperature
of the object at the given phase angle.  The value of 30K for $\Delta T$
was chosen to match the actual brightness temperature
for a rapidly rotating object with an intermediate thermal inertia and a mean
sub-solar illumination direction.

The next step is to compute the monochromatic flux, $F$, incident on
the observatory from the NEA by the Planck blackbody formula, diluted due to
the NEA-observatory distance:
\begin{equation}\label{eq-flux}
F = {D^2 \over 4 \Delta^2} {2 \pi h c^2 \over \lambda^5
( e^{h c \over \lambda k T} - 1 )}
\end{equation}
where $D$ is the diameter of the NEA [km], $\Delta$ is the distance to
the NEA from the observatory [km], $h$ is Planck's constant [erg s], $c$
is the speed of light [cm s$^{-1}$], $\lambda$ is the wavelength of light
[cm], $k$ is Boltzmann's constant [erg deg$^{-1}$], $T$ is the effective
temperature of the NEA [K] taken from Eq.~\ref{eq-tbb}, and $F$ is in
units of [erg cm$^{-2}$ s$^{-1}$].  To get the final detected signal,
we integrated the black-body emission over the detector band pass,
multiply by the area of the telescope, and multiply by
75\% to include the reflectivity of the optics and the quantum
efficiency of the detectors.

\subsection{PSF and trailing losses}

A Sentinel detection comes from seeing a change at a location in the sky.
Such a change is assumed to be a consequence of an object moving across
the field and being detected at different locations with each image.
A detection kernel consisting of a 2x2 group of pixels is used that
will contain flux from a moving object.  A positive detection is thus
one where the detection kernel sees a change in the flux at a given
position by more than 5$\sigma$ above the background signal.

The image formed by a moving object is a trailed PSF.  The trailing is
caused by the apparent motion of the object during the exposure.  Thus,
the signal from the object may not fall in a single detection kernel
if it is moving fast enough.  The trailing fraction is the ratio of 2
(length of detection kernel) divided by the trail length in pixels.
This ratio is capped at unity for very slow moving objects.  The PSF
chosen for the system further reduces the light in the kernel by 55\%.
Combining these two factors yields the fraction of the flux that is
contained within the 2x2 pixel detection kernel.  For the case of a 180
second exposure, any object moving slower than 0.57 degrees/day will incur
minimal trailing losses.  An object moving twice this rate will appear
to be half as bright in a detection kernel for the same emitted flux.

%4.3 arcsec in 180 seconds = 0.0239 arcsec/sec, 86 arcsec/hour, 0.57 deg/day

\subsection{Noise Sources}

There are many sources of statistical fluctuations that limit the
observations such as photon-counting noise from the object, detector
readout noise, detector dark current, thermal background flux from
the optics and telescope structure, and zodiacal light.  Of these, the
zodiacal light, or thermal emission from the dust in the plane of the
solar system, is by far the dominant source of noise.  Constant values
for the adopted readout strategy are read-noise (110 e-/pixel/readout)
and noise from dark-current (600 e-/sec).  Transient noise sources caused
by energetic particles hitting the detector are assumed to be eliminated
by the readout pattern: each 180 second integration is the result of
six 30-sec exposures that are combined in such a way as to eliminate the
transients.  The zodiacal emission is based on a model that integrates
along the line-of-sight from the observatory in the plane of the ecliptic
and over the wavelengths of the system passband.  The three-dimensional
dust cloud is described by an azimuthially symmetric form that decreases
with increasing ecliptic latitude.  This model ignores the fine structure
such as dust bands, the Earth ``ring'', and the small tilt of the cloud
with respect to the ecliptic \citep{kel98}.  The area affected by the
smaller scale structures will locally increase the background but this
is ignored as a small perturbation on the full-sky survey.  The tilt of
the cloud is also ignored for computational convenience.  By ignoring
this tilt, the angular elements of the NEAs and the spacecraft remain
arbitrary.  There is no net effect on the survey sensitivity when used
as a statistical sampling tool.

Given the description of the dust distribution, we estimate the thermal
radiation from the dust along any line of sight with a two-component
model.  The first component computes a detailed line-of-sight integral
through the dust in the plane of the ecliptic.  The second component
is how the flux decreases with increasing ecliptic latitude of the
look direction.  This calculation was developed to quickly compute the
background signal from any vantage point in the inner solar system
near the ecliptic plane.  The computed result tracks the background
signal as a function of the heliocentric distance of the observatory,
the angle from the anti-sun direction, and the angle from the ecliptic.
Modeling a highly inclined observatory orbit would require adding a fourth
dimension to the problem.  This extra complexity was beyond the scope
of this work but we have used an arbitrary factor of 2 reduction in the
zodiacal dust brightness for the one inclined orbit case we considered.
This approximation provides a useful bounding sensitivity test case but
will not represent a rigorous result on the same footing as the orbit
choices that are in the ecliptic.

The line-of-sight integral is estimated by computing the dust density and
temperature along a line {\em in the ecliptic plane} and then determining
the flux across the modeled bandpass.  These integrations are quite slow
and are not practical for embedding directly in the survey simulation
tool.  Instead, a grid of values on these two variables was computed
from which interpolated values are drawn for the simulation.

We created a parameterized expression describing how the zodiacal emission
falls off with increasing ecliptic latitude ($\beta$),
\begin{equation}\label{eq-zodlat}\begin{split}
Z(\beta) = & 1 - 0.919 \biggl\{1-\frac{0.272}{|\sin \beta|}\\
& \left [ 1 - \exp \left ( - \frac{|\beta|}{15.6^{\circ}} - 
\frac{1}{3} \frac{|\beta|^2}{(15.6^{\circ})^2} \right ) \right ] \biggr\}.
\end{split}\end{equation}
Interpolated values from the table were merged with Eq.~\ref{eq-zodlat}
to provide the background flux for any pointing of the observatory
at the location of an object to be detected.  Example values from this
calculation are plotted in Fig.~\ref{fig-zodi}.  This figure shows the
final signal in a detection kernel for two observatory heliocentric
distances (0.7 and 1.0 AU) and at three different ecliptic latitudes
($\beta=$0\mydeg, 30\mydeg\, and 60\mydeg).  The zodiacal background is higher
when the observatory is closer to the sun and looking at lower solar
elongations.  For a similar look angle, the flux is roughly a factor
of three higher at 0.7 AU than it is at 1.0 AU.

\begin{center}
\includegraphics[scale=0.29,trim=80 30 50 30]{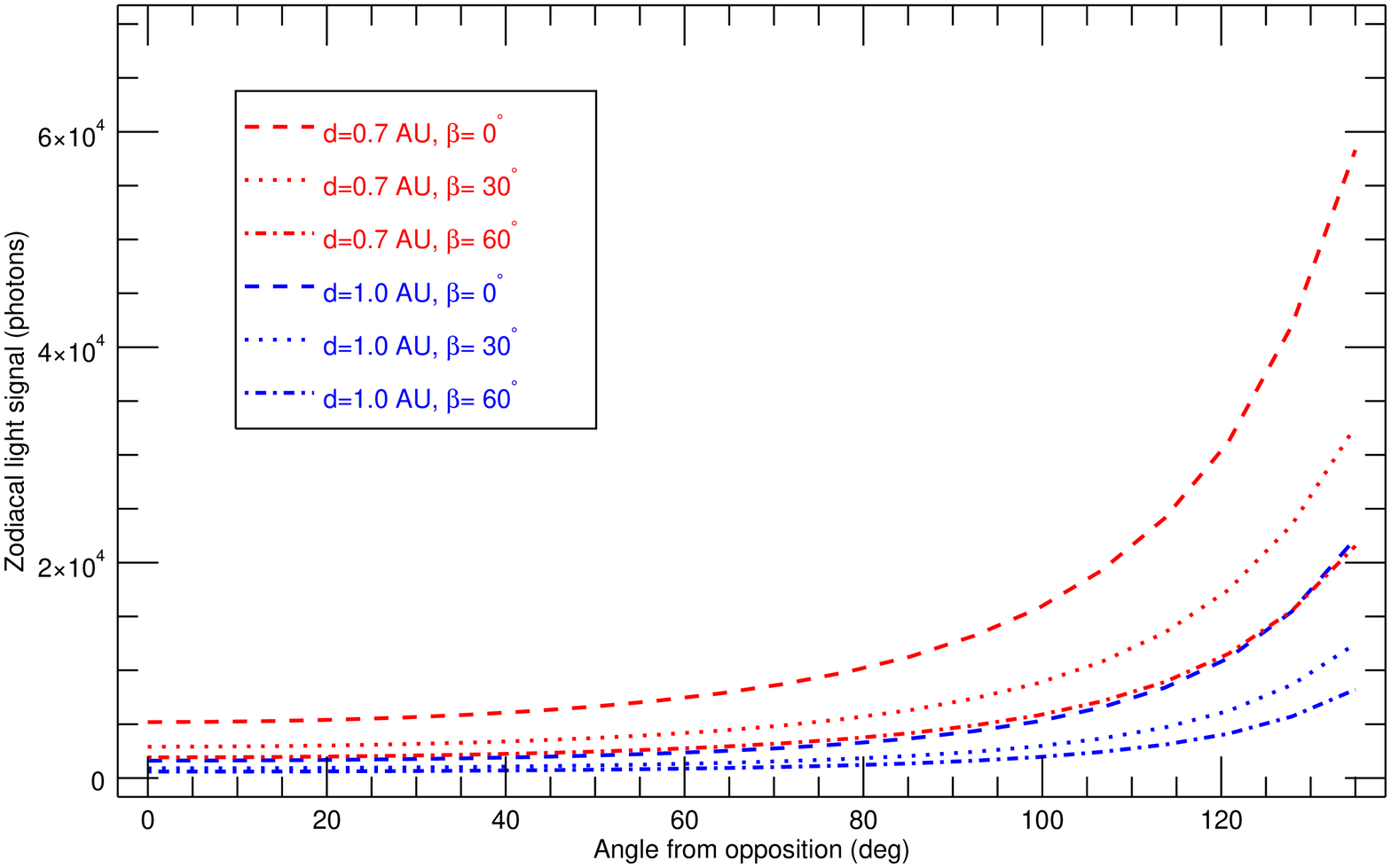}
\figcaption[figzodi.eps]{\label{fig-zodi}\scriptsize
Signal from zodiacal dust in a detection kernel.  The top three (red)
curves show the zodiacal dust signal from a vantage point that is 0.7 AU
from the sun.  The bottom three (blue) curves show the signal that would
be seen at the distance of 1.0 AU.  Pairs of curves with similar linestyles
show the same ecliptic latitude look angle for the two distances.
}
\end{center}

The final SNR for an object is thus the object signal
in the detection kernel divided by the combined noise from all the
noise sources.  To a very good approximation, it is sufficient to use
just the object signal and the shot noise from zodiacal dust
emission as the sole noise terms but we used all of these noise components
in the model since the increase in computation time was negligible.
We also investigated the effects of pixellation on the sampled image with
a numerical experiment that places a source throughout a unit pixel
and mapped this as a function of SNR\null.  At high SNR, it is easy to see
higher signal when the source is centered on the corner of a pixel.  This
geometry maximizes the flux in the 2x2 pixel detection kernel.  At low
SNR, this pixellation pattern disappears.  The result is that regardless
of SNR, a detection is equally likely at all fractional pixel locations.

One final effect was included in the simulations.  Our flux and noise
calculations give a theoretical estimate of the SNR for an observing cycle.
Each opportunity for detection has its own noise, thus an object at SNR=5
will be measured as having an SNR either lower or higher than this value.
We conducted a numerical experiment to measure the odds of 4 successful
detection given an expected SNR\null.  Above SNR=9, the object is always
detected and below SNR=3 the object is never detected with the survey threshold
set to SNR=5.  In between, the probability of 4 successful detections
can be approximated to better than 1\% with the following expression
\begin{equation}\label{eq-tanh}
P_{\rm SNR} = 1 + tanh( ({\rm SNR}-a)/b )/2
\end{equation}
where SNR is the expected signal-to-noise ratio for the source, and $a$
and $b$ are emperically derived factors.  For SNR=5, we used $a=5.99$ and
$b=0.74$.  In our survey model, this expression was used for 3$<$SNR$<$9
at each observing cycle to impose this detection probability on the
synthetic observations.  We compared these results to a simpler case
where the probability is zero if SNR is less than the threshold and one
if higher.  The differences for the two cases is at most a few percentage
points in differential completeness but is systematic in that the largest
effect is at the detection limit.  All of our results include the more
complicated treatment since there is no effective computational penalty
for its inclusion.

\subsection{Fill Factor}

The previous sections indicate if the object will meet or exceed the
signal-to-noise ratio threshold for detection.  The presence of small
gaps between detectors can lead to a non-detection simply because the
object happens to fall in a gap.  For our baseline capabilities, we estimate
that 4\% of the FOV is taken up by gaps that are the boundaries between
individual detectors in the mosaic.  Early on in our work, we used a
simple 96\% probability of detection for each opportunity.  Thus a successful
set of four observations in an observing cycle had to all ``miss the gap.''
We were curious if there were patterns caused by the slightly non-random
velocity vectors of objects lining up with chip gaps that could affect this
simple probabilistic treatment.  A full-fidelity simulation was added where
the object is placed randomly on the FOV for the first of the four
opportunities.  The placement of the other three opportunities relative to the
first was dictated by the orbital motion of the spacecraft and object.
All four of these positions were then checked against a model for the chip
geometry and any position falling in a gap was removed from the detection
list.  The result of this more sophisticated treatment over the entire
course of the survey gave the same answer as the simpler method.  There was
little change in execution time for the software so the more complicated
method was used.

\subsection{Successful Detections and Cataloging}

Combining all of the components in the previous sections decides if an object
is detected or not.  For the four opportunities in each observing cycle,
an object is counted as detected if SNR$\ge$5 given its thermal emission
flux relative to the sky background and detector noise sources and avoids the
inter-chip gaps all four times.  In this analysis we give no credit for
a partial detection (eg., 3 out of 4 avoid gaps).  In practice, these partial
detections would be noted by the spacecraft and transmitted to the ground
because they will still be valid transients.  These data would ultimately
provide some useful measurements despite being ignored for this analysis.
In the end we require a successful detection in at least two observing
cycles (total of eight detections) to consider an object cataloged.

\section{NEA Model}

To compute the results of a survey one must work with a model distribution
of orbital elements of objects.  There are two key components of this
model: the distribution of orbital elements and the absolute number of
objects as a function of size.  These two components are described in
the following two sub-sections.

\subsection{Orbit Model}

The most widely used model for NEA orbit distributions at present is
\citet{bot02}.  This model consists of orbital element bins organized by
semi-major axis, eccentricity, and inclination and a probability of
an object falling within that bin.  Rather than render the distribution
with discrete values of orbital elements, we chose to draw a random set
of elements from the bin based on a uniform probability within the bin.
At the time an orbit is drawn from the distribution, the three angular
elements $\Omega$, $\omega$, and $M$ are also chosen from a uniform
random number between 0 and 2$\pi$.  All orbit calculations for this
survey modeling are done with simple two-body motion.

\subsection{Size Distribution}\label{sect-size}

The survey model includes two methods for incorporating a size
distribution for NEAs.  One direct method is to associate a size with
each orbit where the size is drawn from some approximation of the
size-frequency distribution.  This method becomes more computationally
expensive when working to very small sizes.  The second method is to do
all of the survey calculations based on detection using a single fixed
size.  This type of value is called a differential completeness result
and must be integrated over size to obtain the cumulative survey result
typically used by other investigators.  This result requires multiplying
the computed differential completeness against the differential size
distribution to return the integrated performance.  This method is very
handy in that the size of the object sample can be large enough per size
bin to ensure good statistical sampling of performance.

In addition to the size-frequency distribution, one must also account for
the range of surface albedo seen in the NEA population.  \citet{bot02}
discuss a very simple bi-modal distribution where 9\% of the objects have
an albedo of 0.05 and the rest have an albedo of 0.103 for an absolute
magnitude of $H_V \le 18$.  For fainter objects, 26\% have the lower
albedo and the rest have the higher albedo.  This transition occurs
at roughly $D=1.5$~km.  To support both types of survey calculations, a
set of orbital elements must be generated equal to the total cumulative
number of objects in the size distribution down to the smallest size to
be considered.  For this work we chose to fully model the distribution
down to $D=30$~m.  At smaller sizes we are limited to using differential
calculations.

Our current observational constraints on the size distribution have a
photometric basis (apparent brightness) and are most strongly guided
by reflected-light optical surveys.  Thus, what we really have is a
brightness distribution.  Such distributions need to be converted to
physical size before modeling can begin.  It was beyond the scope of this
work to properly convert from a brightness distribution to a true size
distribution given the relatively poor state of knowledge of the actual
albedo distribution for sizes below a few hundred meters in diameter.

The size distribution of \citet{bos15} was used to support this survey
model and a sample of 3.5 million orbits was generated to support
a cumulative calculation for $D>30$~m.  The value drawn from the
distribution was the H$_V$ magnitude according to the bounded probability
function.  Once a value for H$_V$ was chosen, the albedo was selected
from the two discrete values in accordance with the probability of those
albedos given the size of the object.

\subsection{Main Belt Asteroids}

An NEA survey will also see main-belt asteroids (MBAs) during
its operation.  It is useful to estimate the expected number
of MBA observations from a survey.  Rather than use a synthetic
distribution as with NEAs, we chose to use orbits and properties
from known asteroids.  Owing to their greater distance, there is
no expectation of getting to the same small size regime as for NEAs.
The known MBAs are complete enough to estimate the rate of re-detection
for this purpose.  These objects could also be used as calibrators for
a survey for testing sensitivity and confirming any debiasing methods
that would be applied to the NEA data.  A secondary database table
was built from all of the known asteroids using the astorb database
developed by E. Bowell and now maintained by L. Wasserman at Lowell
Observatory\footnote{\url{ftp://ftp.lowell.edu/pub/elgb/astorb.dat}}.
This database includes assumptions or data regarding geometric
albedo and phase function and this was used as is to compute physical
sizes.  The objects used were all those with multi-apparition orbits.
The database for these tests was generated on Nov. 2013 at which time
there were 536,261 objects.

\subsection{Object and Orbit Sampling}

The set of NEA orbits can be used either as a complete sample or they
can be statistically sampled.  To use the data as a complete sample,
the entire list is run through the survey simulator with some size
cutoff.  This method permits directly sampling the survey properties
on a cumulative size distribution with an embedded albedo distribution.
Unfortunately, using a cumulative size distribution requires testing a
very large number of orbits and the execution time can be prohibitively
long.

An alternative is to use the orbits as a set from which to pull a
statistically meaningful sample.  In this way, a much smaller number
of orbits can be used with substantial reduction in the run-time while
ensuring good sampling at all sizes.  The procedure used is to set a
fixed size and albedo to a sub-sample of orbits and measure the survey
properties.  By running the simulator over a small number of discrete
sizes and albedos, the differential completeness properties of the
survey are measured.  The quantity measured is the fraction observed
as a function of time.  The results presented here are based on 20,000
randomly chosen orbits from the full orbit sample.  This sample size
provides reasonably good statistics for the sizes considered here but
some calculations at the smallest target sizes require an increase to
counteract the low completeness values.

\subsection{Sampling Other Object Types}\label{sect-samp}

The nominal set of NEA orbits from the Bottke distribution is a very
diverse set of orbital elements.  The only criterion applied for these
objects is that their perihelion distance must be less than 1.3 AU ---
corresponding to a commonly accepted definition for a near-Earth asteroid.
This set of objects is considered to be the baseline reference set due
to the current Congressional challenge to NASA\footnote{G. E. Brown,
Jr., Near-Earth Object Survey Act. NASA Authorization Act of 2005
(Public Law No. 109-155), referred to hereafter as the GEB survey
mandate.} and the scientific community\footnote{Near Earth Object
Science Definition Team, 2003. Study to Determine the Feasibility
of Extending the Search for Near-Earth Objects to Smaller Limiting
Diameters. \url{http://neo.jpl.nasa.gov/neo/neoreport030825.pdf}, accessed
9/14/2015.}.  The language at the time is most readily interpreted as
referring to NEAs.

There are other subsets of objects worthy of consideration as well.
These objects are contained within the set referred to as NEAs and
their physical properties are generally considered to be the same.
The most common subset is dubbed potentially hazardous asteroids (PHAs).
These bodies are in orbits with minimum orbital intersection distances
(MOIDs) with respect to the Earth $\le 0.05$~AU.  To support working with
this subset, the MOID for each sampled object from the Bottke distribution
was computed and saved to the database.  Using the sampled set of orbits,
the ratio of the number of PHAs to NEAs down to a given size is 0.202.
Note that it is common to impose a lower size bound to objects labeled
as PHAs.  We do not include this cutoff in the analysis and only use
the orbital properties to identify PHAs.  To complete the usual set of
sub-populations we also considered, Apollos ($q \le 1.0167$, $a \ge
1$), Atens ($q < 1$, $Q \ge 0.983$), Earth-Crossing Asteroids (ECAs,
Apollos$+$Atens), Amors ($a \ge 1$, $q > 1.0167$), and Atiras ($a < 1$,
$Q < 0.983$), where $a$ is the semi-major axis of the orbit, $q$ is the
perihelion distance, and $Q$ is the aphelion distance and all of these
values are given in AU.

% In database:
%
% select count(*) from orb02,phys02
% where orb02.id=phys02.id and semi*(1-ecc)<1.30 and diam>0.015;
%
% This is used to get the ratio of NEA to PHA's, not for the total number
%    The value is 0.202 = N_PHA/N_NEA

Another set of objects are those of interest as human exploration targets.
Objects in these orbits are accessible to direct exploration via easier
low-energy launches.  Paul Chodas (personal communication) gave the
orbit element range of $0.85 < a < 1.25$, $e<0.17$, and $i<6^{\circ}$
for orbits with $V_{\infty} \le 2$ km/s that meet these requirements.
We use this approximation of an orbital-element selection criterion
for these objects due to the lack of actual knowledge of $V_{\infty}$
within our simulator.  This different samples of objects can be tested by
modifying the database query to the appropriate set of orbital elements.
The fraction of the NEA population that satisfy these constraints is
rather small so a special sample was created to ensure the detection
statistics would be accurate.  This population is referred to in this
work as the Asteroid Robotic Redirect Mission (ARRM) sample.

To better study objects that pose a more direct threat to Earth,
we used the sample of virtual impactors (VIMP) that were published by
\citet{che04} and \citet{ver09}.  These orbits are drawn from the Bottke
distribution but are determined to be orbits that result in an impact on
the Earth with a uniform probability of impact over a 100 year time span.
In keeping with this work, we have also adopted 100 years as a time-span
of interest because it matches a human lifespan while also being a
duration over which deterministic impact calculations can be made with
sufficient supporting astrometric information.  This also happens to match
what the JPL Sentry system\footnote{\url{http://neo.jpl.nasa.gov/risk/}}
uses.  This sample of impactor orbits contains 10,000 entries and require
special care when used for very small sizes where the survey completeness
is very low.  For larger sizes this permits comparing mission profiles
to evaluate their effectiveness in finding objects that are impactors
within 100 years of the end of the survey.

All of these sub-populations must be scaled before computing the absolute
numbers of detections and cumulative completeness.  Table~\ref{tbl-norm}
provides the adopted scaling fractions for the sub-populations
as represented in the \citet{bot02} distribution compared to the
full NEA population.  These values are multiplied against the NEA
size distribution curve from \citet{bos15} to get the absolute size
distribution curve for a sub-population.  The second column lists the
fraction of that sub-population compared to the total NEA population and
is assumed to be independent of size.  Using an impact probability for
ECAs of 2.8$\times$10$^{-9}$ per year (W. Bottke, private communication)
multiplied by the ECA fraction and by 100 gives an estimate of the total
impacting flux on the Earth in a 100 year period.  This last case is
listed as Impactors but is also referred to as ``VIMP.''  The entries in
the table are sorted by decreasing population.  Note that when we model
all of these populations, we consider the entire population and do not
consider that some fraction of these are already known.

% full res value is 2.7978e-9 from Bottke

\begin{deluxetable}{cccl}
\tablecaption{Population Normalizations\label{tbl-norm}}
\tablewidth{0pt}
\tablehead{
\colhead{Type}&
\colhead{N/N$_{NEA}$}&
\colhead{Orbit Constraint}
}
\startdata
NEA&    1.&       $q<1.3$\\
ECA&    0.683&    Apollo$+$Aten\\
Apollo& 0.623&    $q \le 1.0167$, $a \ge 1$\\
Amor&   0.295&    $a \ge 1$, $q > 1.0167$\\
PHA&    0.202&    $q<1.3$, MOID $\le 0.05$\\
Aten&   0.0598&   $q < 1$, $Q \ge 0.983$\\
Atira&  0.0217&   $a < 1$, $Q < 0.983$\\
ARRM&   0.000348& $0.85 < a < 1.25$, $e<0.17$, $i<6^{\circ}$\\
Impactors& 1.9$\times$10$^{-7}$& Earth impactors\\
\enddata
\tablecomments{\scriptsize
The fraction for impactors gives the number per century.
}
\end{deluxetable}

\subsection{Field of Regard}

The area in the sky that will be surveyed is called the field of regard
(FOR).  This region is defined in the baseline mission as running from
80\mydeg\ solar elongation to the anti-sun direction and over the full
range of ecliptic latitude.  The dwell time per field and the field of
view (FOV) of the camera require 28 days to cover the entire FOR in the
baseline mission.  This cadence includes all slew and settling overhead
plus operating margin and returns four observations of every location
in the FOR per cycle.

In the simulator, the FOR can be varied by setting the range of ecliptic
latitude and solar elongation to cover.  Changing these ranges will
alter the time required to cover the FOR and thus change the time between
successive observing cycles.  The revised observing cycle time is scaled
from the baseline 28 days by the ratio of the sky area coverage of the
revised FOR to the sky area covered by the baseline FOR.  We require
that the arc-length for the observations be at least 28 days to tag an
object as being cataloged.

\section{Detection Efficiencies}

In the simulator, each object is tested at each possible detection time
for falling in a gap between detectors.  The relative motion of the
object between the individual observations is tracked but this set of
opportunities is randomly placed on the FOV.  If the object is bright
enough to be detected, is in the FOR, and avoids the inter-chip gaps all
four times it is counted as being detected in that observing cycle.

\subsection{Differential Detection Efficiencies}

The primary results from a simulation of the baseline Sentinel survey
are summarized in Table~\ref{tbl-eff}.  The absolute magnitudes ($H_V$)
and the total number N$_{\rm NEA}$ come directly from \citet{bos15}.
The tabulated diameter is the result of a simple conversion to size using
a fixed geometric albedo of 14\%.  The values labeled with ``N'' give
the current estimate of object greater than or equal to the tabulated
diameter.  The values labeled with ``$d$'' represent the completion
percentage for the tabulated size -- the differential completeness as
described in section~\ref{sect-size}.  The values labeled with ``$c$''
represent the cumulative completeness.  The values labeled with ``T'' are
estimates of the total number of objects to be cataloged by the survey.
The values labeled with ``NEA'' are drawn directly from the \citep{bot02}
distribution subject to the perihelion distance constraint ($a \le 1.3$).
The values labeled with ``PHA'' are drawn from the NEA distribution but
also subject to the constraint of MOID$\le 0.05$.  The values labeled with
``VIMP'' represent the impactor population.  The values with parentheses
are provided in shortened form for exponential notation where the power
of ten multiplier is given in parentheses.  It's worth considering that
the NEAs have a much broader range of orbital properties compared to PHAs
and PHAs are much broader than VIMPs.  NEAs include some objects that
never get that close to the Earth (for example, $i=80^{\circ}$, $e=0.9$,
and $q=1.3$).  Also, these relatively extreme orbits (particularly those
with high eccentricity) have longer periods than the mission duration
and some never get close enough to detect at all.  A major result of
this simulation is that our survey design reaches a significantly higher
completion level for those objects that are a direct threat to the Earth
than for the much broader and less threatening NEA population.

%
%Checked against Boslough and 14%
%
\begin{deluxetable}{cccccccccccccc}
\rotate
\tabletypesize{\scriptsize}
\tablecaption{Completeness -- Baseline Survey\label{tbl-eff}}
\tablewidth{0pt}
\tablehead{
\colhead{$H_V$} &
\colhead{D} &
\colhead{N$_{\rm NEA}$} &
\colhead{N$_{\rm PHA}$} &
\colhead{N$_{\rm VIMP}$} &
\colhead{$d_{\rm NEA}$} &
\colhead{$d_{\rm PHA}$} &
\colhead{$d_{\rm VIMP}$} &
\colhead{$c_{\rm NEA}$} &
\colhead{$c_{\rm PHA}$} &
\colhead{$c_{\rm VIMP}$} &
\colhead{T$_{\rm NEA}$} &
\colhead{T$_{\rm PHA}$} &
\colhead{T$_{\rm VIMP}$}\cr
& (km)& & & & \%& \%& \%& \%& \%& \%&
}
\startdata
26.9& 0.015& 34,600,000& 7,000,000& 6.6e$+$00&   0.3&  0.4&   2.5&   2.1&  2.8&        8& 714,000& 198,000& 5.3e$-$1\cr
26.3& 0.020& 15,800,000& 3,200,000& 3.0e$+$00&   1.1&  1.4&   6.0&   4.1&  5.6&       14& 647,000& 180,000& 4.3e$-$1\cr
25.4& 0.030&  4,790,000&   970,000& 9.2e$-$01&   4.5&  6.4&    18&    10&   14&       30& 493,000& 139,000& 2.8e$-$1\cr
23.9& 0.060&    513,000&   104,000& 9.8e$-$02&    24&   34&    61&    39&   48&       70& 198,000&  49,400& 6.9e$-$2\cr
22.8& 0.100&    142,000&    28,700& 2.7e$-$02&    52&   60&    80&    69&   74&       88&  98,000&  21,200& 2.4e$-$2\cr
22.1& 0.140&     75,800&    15,300& 1.4e$-$02&    69&   73&    88&    83&   85&       94&  63,000&  13,100& 1.4e$-$2\cr
21.3& 0.200&     46,100&     9,300& 8.8e$-$03&    82&   84&    94&    91&   92&       97&  42,000&   8,590& 8.6e$-$3\cr
19.9& 0.380&     19,600&     4,000& 3.8e$-$03&    96&   96&   $>$99&  98&   98&    $>$99&  19,200&   3,900& 3.7e$-$3\cr
17.8& 1.000&      4,440&       898& 8.5e$-$04&  $>$99& $>$99& $>$99& $>$99& $>$99& $>$99&   4,410&   895&   8.5e$-$4\cr
16.9& 1.514&      2,230&       450& 4.3e$-$04&  $>$99& $>$99& $>$99& $>$99& $>$99& $>$99&   2,210&   448&   4.2e$-$4\cr
15.9& 2.399&        951&       192& 1.8e$-$04&  $>$99& $>$99& $>$99& $>$99& $>$99& $>$99&     939&   190&   1.8e$-$4\cr
\enddata
\tablecomments{\scriptsize
Completeness values are given as a function of size and absolute brightness.
The simulation and conversion from $D$ to $H_V$ was based on a fixed
geometric albedo of 14\%.  The values labeled with ``N'' are absolute
cumulative numbers for the different populations given by the subscript.
The differential (single-size) completeness values are labled with ``$d$''
and the cumulative completeness values are labeled with ``$c$''.  The columns
labeled with ``T'' give the total cumulative number of objects detected.
}
\end{deluxetable}

%IDL> distfun,'v_neo_06.5_5.0_050_0.14_cum.dat',/nosave
%IDL> distfun,'v_pho_06.5_5.0_050_0.14_cum.dat',/nosave
%IDL> distfun,'v_vimp_06.5_5.0_050_0.14_cum.dat',/nosave

\subsection{Cumulative Detection Efficiencies}

An important outcome of the survey simulations is to quantify the
absolute number of objects that will be detected.  Converting from
the differential values ($d$) to the cumulative values ($c$) requires
integrating the product of $d$ and the derivative of the Harris
distribution.  Table~\ref{tbl-eff} shows these results.  The integral
is performed over $H_V$ but a corresponding size ($D$) is shown for
a fixed 14\% geometric albedo.  The cumulative number of NEAs down
to the tabulated size is given in the column labeled N$_{\rm NEA}$.
These values directly replicate the Harris distribution.  The next
column gives the total number of PHAs and is based on a fixed fraction
determined from the orbit distribution itself.  In the \citet{bot02}
distribution, 20.2\% of the NEAs are PHAs (see Table~\ref{tbl-norm}).
The size dependent cumulative completeness values and the total number
of objects detected ($T$) are thus based on this calibration sequence.
The completeness for the baseline survey and $D \ge 140$m is 85\% for
NEAs, 87\% for PHAs, and 95\% for VIMPs.  Clearly, the completeness of
the survey diminishes with decreasing size.  The 50\% efficiency point
lies between 60-100m for NEAs and PHAs but lies between 30-60m for VIMPs.
Even though the differential completeness falls off dramatically at very
small sizes, the total number of small objects that will be detected is
substantial.  Such a survey will be very well suited to a substantial
refinement or confirmation of the size distribution over the range of
sizes that represent significant threats to the Earth.  Objects smaller
than those tabulated are very unlikely to cause major damage upon impact.

The issue of albedo distribution presents a significant challenge
when providing any cumulative results that indicate both size and
absolute magnitude.  It has become common to use 14\% geometric albedo to
convert between these two quantities but this is only an approximation.
Showing results for a plot of cumulative detections really should
have two separate curves, one for detections vs. size and one for
detections vs. absolute magnitude.  In the case of modeling optical
surveys, this is an extremely important point.  The detected flux for
an object depends linearly on the albedo.  As shown by \citet{mai11},
the NEA albedos are far from uniform with a range from 2-3\% up to as
high as 35\%.  For a given size, the observed brightness of an object
is directly proportional to albedo making this a critical factor for
proper modeling of optical surveys.  The situation is not as extreme
for thermal infrared detection surveys since the flux dependency on
albedo is very weak in that it varies as the fourth root of albedo.
Over this range of expected albedos for NEAs, the variation in detected
IR flux is at most a 10\% effect \citep{spe89}.  For this reason, we
chose to simplify our modeling by using a single albedo rather than a
distribution to make it easier to understand the results versus both
albedo and absolute magnitude, thus facilitating comparisons with past
survey results.  A more sophisticated treatment will change the absolute
numbers by a small amount but the basic results we present will remain.
In truth, the real distribution of NEA properties is more complex and the
mapping of absolute magnitude to size is more complex.  For the purposes
of our survey design investigations, a simple mapping using a uniform 14\%
albedo was sufficient.  It is clear to us, however, that there is a lot
of merit in complementary surveys.  Having a pair of surveys running,
one in the optical and one in the infrared, is an excellent hedge against
systematic detection biases when the goal is to find all the objects.
These considerations are less important if the goal is merely to get enough
objects to measure of the distribution after removing biases.  In the case
of Sentinel, the primary goal is to find objects and extracting a size
distribution is of secondary importance.

\subsection{Synodic Period Considerations}

For a given simulator run, the synodic period was calculated for each
object based on the observatory orbit.  The sidereal periods are also
computed.  Any object whose sidereal or synodic period is longer than
the mission duration could remain undetectable for the entire survey.
This fraction of objects is really an upper limit to the number that can
escape detection.  For example, an object could be large enough to be
detectable over the entire orbit and the synodic period would not matter.
There are other intermediate cases where the object is detectible over
enough of the synodic period that again it does not matter.  Also, an
object in a low-eccentricity orbit might always be detectable whereas
the same size object in a high-eccentricity orbit would be out of range
for most of its orbit.  Regardless of this complexity, the upper limit
provides an indication of the fraction of the population that might never
be seen.  For the baseline mission, NEAs contain 2.7\% with long synodic
or sidereal periods, PHAs contain 1.6\% that are too long, and VIMPs
have only 1.2\% in this category.  Essentially all of the long-period
VIMPs are long by virtue of their synodic periods with less than 0.01\%
having long sidereal periods.  The complication of long synodic period
objects appears to be a small component of the challenge of surveying
the NEA population, at least for size ranges with low completion rates.

\subsection{Modeling uncertainties}

Our method of using differential completeness calculations was employed to
minimize the numerical errors in the survey modeling.  The counting errors
in the model for a single size can be easily characterized and controlled
and depend only on the number of objects in the sample.  We standardized
our calculations on 20,000 synthetic objects.  This sample size results
in completeness values good to $\le$0.1\% down to $D=30$m.

The uncertainties in the final cumulative completeness results is
thus dominated by our limited knowledge of the actual properties
of the NEA population.  None of the distributions of size, albedo,
thermal properties, pole position, and rotation periods are known
perfectly well.  We have used our own approximations for these values
and different choices for these distributions will change the results
by more than the sampling uncertainties in our model.  For this reason,
we do not delve more deeply into the modeling uncertainties as this really
is a long-term research problem into the actual properties of NEAs and
is beyond the scope of this work.

\section{Alternate Survey Strategies}

The survey simulator allows the exploration of survey designs other than
the baseline mission concept.  To speed up the exploration of parameter
space, differential efficiencies at $D=140$~m and $D=30$~m are computed.
From the cumulative results, the $D=140$~m efficiency should be around
85\% to ensure a cumulative completeness of 90\%.  This size is thus used
to estimate if a given strategy meets the overall mission objectives as
well as being an overall improvement or not.  A second case of $D=30$m,
significant to Planetary Defense, is also computed to
see if there are options that might excel at picking up smaller objects
while not compromising the overall goal of getting a more complete census
for $D \ge 140$~m.

The simulator permits altering many aspects of the survey: mission
duration, telescope aperture, sky coverage (FOR), field-of-view, exposure
time, detection threshold, and spacecraft orbit.  Different asteroid
orbit distribution models can also be used as they become available.
The following sub-sections summarize the basic results of these survey
variations compared to the baseline.  The results from the analyses in
this section are summarized in Table~\ref{tbl-trade}.  The first case
shown is the reference values for the baseline mission (V).  The numbers
shown are all given as percentages of differential completeness for the
indicated diameter in meters for three classes of objects: NEAs, PHAs, and
impactors.  The columns labeled ``syn'' indicate the percentage of objects
with synodic periods longer than the 6.5-year baseline mission duration.

\begin{deluxetable}{ccccccccccl}
\tablecaption{Completeness Trades\label{tbl-trade}}
\tablewidth{0pt}
\tablehead{
Case & \multispan3{\hfil NEA\hfil}&
\multispan3{\hfil PHA\hfil}& \multispan3{\hfil Impactors\hfil}& Notes\\
&
\colhead{$d_{140}$}&
\colhead{$d_{30}$}&
\colhead{syn}&
\colhead{$d_{140}$}&
\colhead{$d_{30}$}&
\colhead{syn}&
\colhead{$d_{140}$}&
\colhead{$d_{30}$}&
\colhead{syn}&
}
\startdata
  V& 69& 4.5& 2.6& 73& 6.4& 1.7& 88& 18& 1.2& Baseline mission\\
 VE& 61& 3.1& 3.6& 65& 6.0& 3.3& 81& 21& 5.5& No gravity assist\\
  E& 51& 2.5& 7.2& 56& 4.5& 7.6& 67& 17&  16& Earth-similar orbit\\
  N& 59& 0.9& 7.1& 66& 2.9& 7.4& 76& 13&  16& NEOcam FOR near Earth\\
V10& 81& 6.3& 0.6& 84&  10& 0.3& 94& 25& 0.7& Case V for 10 years\\
 VX& 86&  10& 0.7& 89&  16& 0.4& 96& 39& 0.7& Enhanced Case V\\
\enddata
\vskip-20pt
\tablecomments{\scriptsize
The case shown is an abbreviation for the orbit or mission option studied
(see text for details).  The differential completeness ($d$) are shown for
three sets of objects and the sub-script refers to the size, in meters,
modeled.  The column labeled ``syn'' indicates the percentage of the sample
that has either a synodic period or orbit period greater than the 6.5 year
the survey duration.
}
\end{deluxetable}

% V   v_neo_06.5_5.0_050_0.14_cum.dat
%     v_pho_06.5_5.0_050_0.14_cum.dat
%     v_vimp_06.5_5.0_050_0.14_cum.dat
% VE  ve_neo_06.5_5.0_050_0.14_cum.dat
%     ve_pho_06.5_5.0_050_0.14_cum.dat
%     ve_vimp_06.5_5.0_050_0.14_cum.dat
% E   e_neo_06.5_5.0_050_0.14_cum.dat
%     e_pho_06.5_5.0_050_0.14_cum.dat
%     e_vimp_06.5_5.0_050_0.14_cum.dat
% N   n_neo_06.5_5.0_050_0.14_cum.dat
%     n_pho_06.5_5.0_050_0.14_cum.dat
%     n_vimp_06.5_5.0_050_0.14_cum.dat
%V10
%survey,'NEO',0.140,'scorbit.dat',-90,90, 80,180,10.0,/nodisplay
%survey,'NEO',0.030,'scorbit.dat',-90,90, 80,180,10.0,/nodisplay
%survey,'PHO',0.140,'scorbit.dat',-90,90, 80,180,10.0,/nodisplay
%survey,'PHO',0.030,'scorbit.dat',-90,90, 80,180,10.0,/nodisplay
%survey,'VIMP',0.140,'scorbit.dat',-90,90, 80,180,10.0,/nodisplay
%survey,'VIMP',0.030,'scorbit.dat',-90,90, 80,180,10.0,/nodisplay
%VX
%survey,'NEO',0.140,'scorbit.dat',-90,90, 80,180,10.0,apfac=65.0/50.0,/nodisplay
%survey,'NEO',0.030,'scorbit.dat',-90,90, 80,180,10.0,apfac=65.0/50.0,/nodisplay
%survey,'PHO',0.140,'scorbit.dat',-90,90, 80,180,10.0,apfac=65.0/50.0,/nodisplay
%survey,'PHO',0.030,'scorbit.dat',-90,90, 80,180,10.0,apfac=65.0/50.0,/nodisplay
%survey,'VIMP',0.140,'scorbit.dat',-90,90, 80,180,10.0,apfac=65.0/50.0,/nodisplay
%survey,'VIMP',0.030,'scorbit.dat',-90,90, 80,180,10.0,apfac=65.0/50.0,/nodisplay

\subsection{Alternate Orbits}

There are many orbital parameters that could be varied to investigate
their potential for improved performance over the baseline Sentinel
orbit.  Interior orbits work well for providing a geometry very
different from an Earth-based vantage point.  As demonstrated earlier,
the zodiacal background is higher with decreasing heliocentric distance
for the observatory and is lower when working near the opposition
region. These are two examples where the outcome of the change in orbit
is not immediately obvious and the simulator can provide useful guiding
information.  

\subsubsection{No Venus Gravity Assist}

The baseline mission uses a Venus gravity assist to reduce the aphelion
distance.  This maneuver requires no fuel but it does impose timing
constraints on the launch window.  Without the gravity assist one can
choose, using only the Earth-departure launch vehicle, a nearly arbitrary
perihelion distance with aphelion at 1 AU, or, an arbitrary aphelion
distance with perihelion at 1 AU ($a$=0.8 AU, $e$=0.25).  Orbits that
extend beyond 1 AU were not investigated since that vantage point is
severely handicapped for finding objects when they are interior to
the Earth.

To investigate this option we chose an orbit with the same perihelion as
the baseline orbit but an eccentricity that puts aphelion at 1 AU\null.
Since the small end of the size range is limited by range to the target,
this might have a chance to improve finding smaller objects that are in
more Earth-like orbits.  As shown in Table~\ref{tbl-trade}, the survey
performance is worse for $D=140$~m regardless of the population tracked.
The performance is also worse for $D=30$~m for NEAs and PHAs but is
better for impactors.  This option is denoted as mission option VE for
the orbit carrying the observatory between the orbit of Venus and Earth.
All other properties of Sentinel are kept the same.

Unfortunately, limitations in the current treatment of the zodiacal
background prevent testing orbital options for a spacecraft that is
significantly out of the ecliptic plane.  A highly inclined orbit
might be useful so that objects in the ecliptic are not seen against
the strongest zodiacal emission in the plane.  However, as seen in
Fig.~\ref{fig-zodi}, the best this option can provide is roughly a factor
of 2-3 reduction in background.  As a proxy for this exploration, a
case was investigated where the background flux is arbitrarily reduced
in combination with a high-inclination orbit.  The performance in this
orbit was not markedly different from the performance of other cases
studied and those model results are not discussed further.

\subsubsection{Earth-similar Orbit}

An even lower launch energy option is an orbit that puts it near the
Earth.  For the simulator it makes no difference if the observatory
is in orbit around the Earth, at the L1 or L2 Lagrange points, or in an
Earth-trailing or Earth-leading orbit.  In reality, these options do place
additional constraints on the accessible region of sky and how a survey
might be carried out but these effects were ignored for this analysis.

Using an Earth-like orbit degrades the survey performance even more
than the previous no-gravity-assist case (see Table~\ref{tbl-trade}).
However, the degradation is less at the small end of the size range.
The penalty induced by long synodic period objects is noticeably worse
for this option and is worst of all for the impactors.  In this case,
15\% of the VIMP sample has a synodic period longer than the 6.5 year
survey.

Note that a space observatory located near the Earth will be observing
a very similar subset of the population as is accessible to ground-based
observatories at the same time -- an asset if the survey goal is to get
complementary thermo-physical observations.  However, this orbit option
is less desirable if the goal is to discover new objects as quickly
as possible.  This option is denoted as mission option E since the orbit
is essentially the same as Earth.  All other properties of Sentinel are
kept the same.

\subsubsection{Other Options}

Also included in Table~\ref{tbl-trade} is a case of an Earth-like orbit
with a FOR similar to NEOCam as described in \citet{mai15a} and denoted
as case N\null.  The performance of case N is better than case E for NEAs
and PHAs but worse for impactors.  Case V10 is the same as the baseline
mission but for a 10-year survey duration.  Running the survey longer
clearly improves the performance in all categories.  Similar improvements
would be seen for any orbit option.  One final case, VX, is the baseline
mission operating for 10 years with a 65-cm primary mirror.  The FOV
and pixel sampling relative to the diffraction limit is assumed to be
the same as the baseline mission.  All other mission parameters remain
the same.  These last two cases are clearly higher-performance surveys
but the changes to the mission would add significant costs to the mission.

\subsection{Sky-plane Detection Maps}

Understanding the result of a given choice of FOR is enhanced by looking
at sky-plane maps where objects can be detected.  To get this, the
positions of all detectable objects are saved at each time step during
a simulated survey.  This information ignores the FOR within the survey
run but includes source brightness, trailing losses, and background
brightness.  The pattern is collected in a coordinate system measured
in ecliptic latitude and angle from the sun.  This view will converge
on a long-term average result.  To get these views, the survey was run
as long as needed to build up meaningful statistics.  For clarity, the
region within 25\mydeg\ of the Sun is suppressed.  These results are much
more informative when computed for a single size range at targeted sizes.

Figure \ref{fig-neo-v} shows sky-plane detection maps for NEAs at a
few strategic sizes for the baseline Sentinel mission.  These images
show relative likelihoods of detection for a single size of object
as indicated with the most likely location being normalized to unity.
The objects are largely concentrated within $\pm$30\mydeg of the ecliptic.
The contributions near the ecliptic poles never drop to zero but
\begin{center}
\includegraphics[scale=0.47,trim=80 20 50 30]{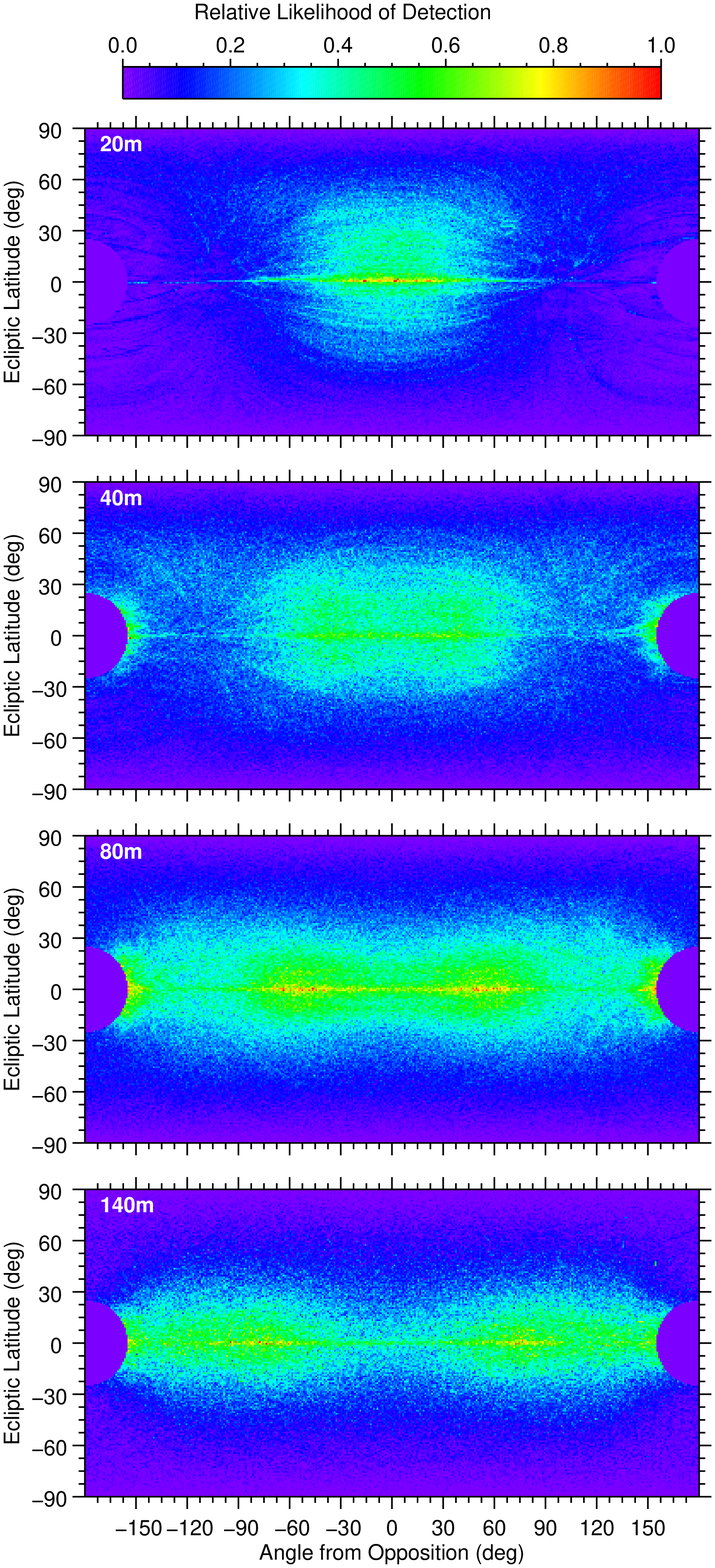}
\figcaption[fig-neo-v.eps]{\label{fig-neo-v}\scriptsize
Detectability map for Near-Earth Asteroids with the nominal Sentinel
design (V) for different target sizes.  Each panel shows where in the sky
a target of the given size can be detected.  The top panel is for the
smallest objects with $D=20$~m and the bottom panel is for $D=140$~m.  Each
map is normalized to 1 at the most likely location.  Points with 25\mydeg\
of the Sun are set to zero.  This sequence clearly shows a transition from
favoring quadrature for detecting larger objects to opposition for detecting
small objects.
}
\end{center}
is much less than along the ecliptic.  Most interesting is the trend with
object size.  The D=140m map at the bottom of the figure shows that the
regions near quadrature are favored for object
\begin{center}
\includegraphics[scale=0.47,trim=80 20 50 30]{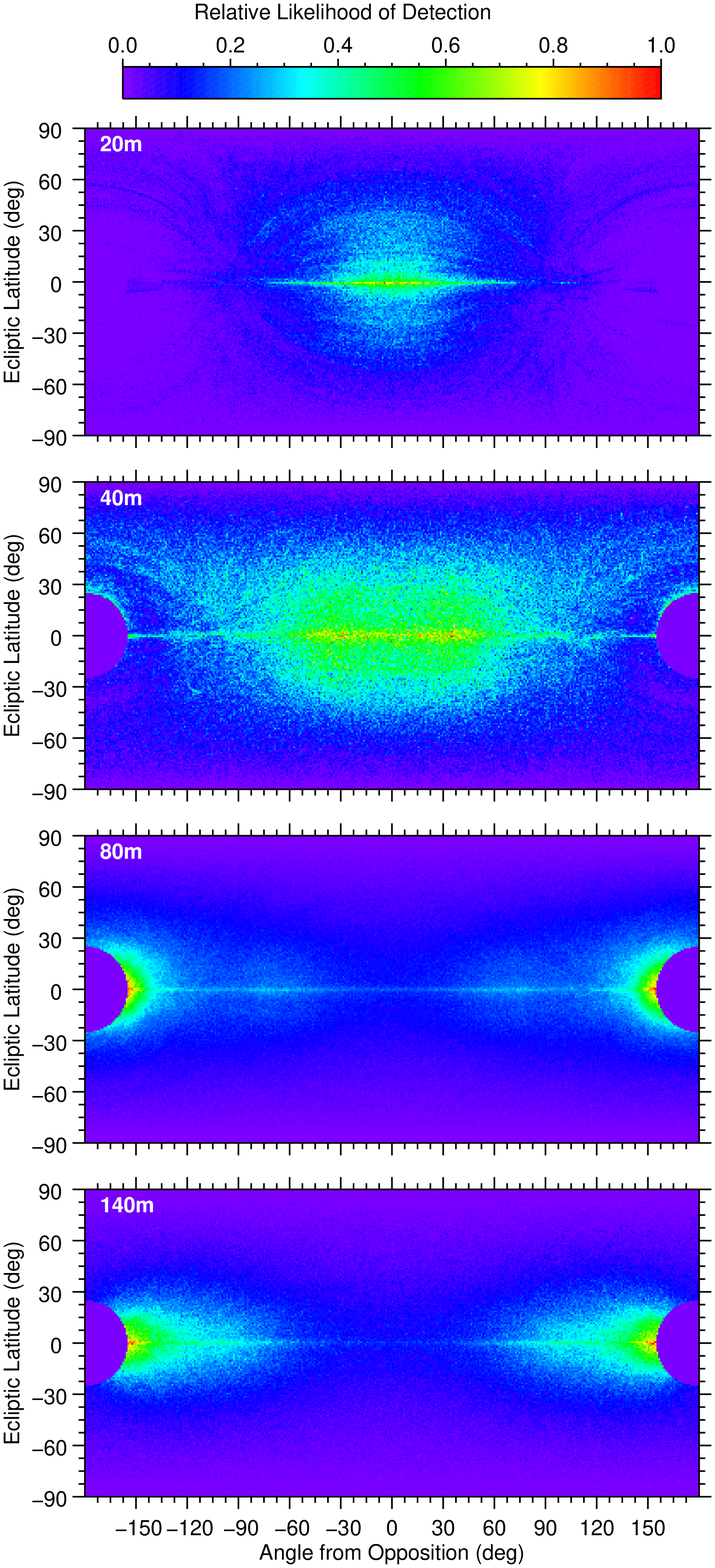}
\figcaption[fig-neo-e.eps]{\label{fig-neo-e}\scriptsize
Detectability map for Near-Earth Asteroids with the mission option E
for different target sizes.  Each panel shows where in the sky
a target of the given size can be detected.  The top panel is for the
smallest objects with $D=20$~m and the bottom panel is for $D=140$~m.  Each
map is normalized to 1 at the most likely location.  Points with 25\mydeg\
of the Sun are set to zero.  At this heliocentric distance, detections are
more likely at lower solar elongation angles as the object size increases.
}
\end{center}
detection.  As the object
size decreases there is a marked change in the pattern to where the
opposition region is strongly preferred once the size drops to $D=20$~m.
This trend has been reported
\begin{center}
\includegraphics[scale=0.47,trim=80 20 50 30]{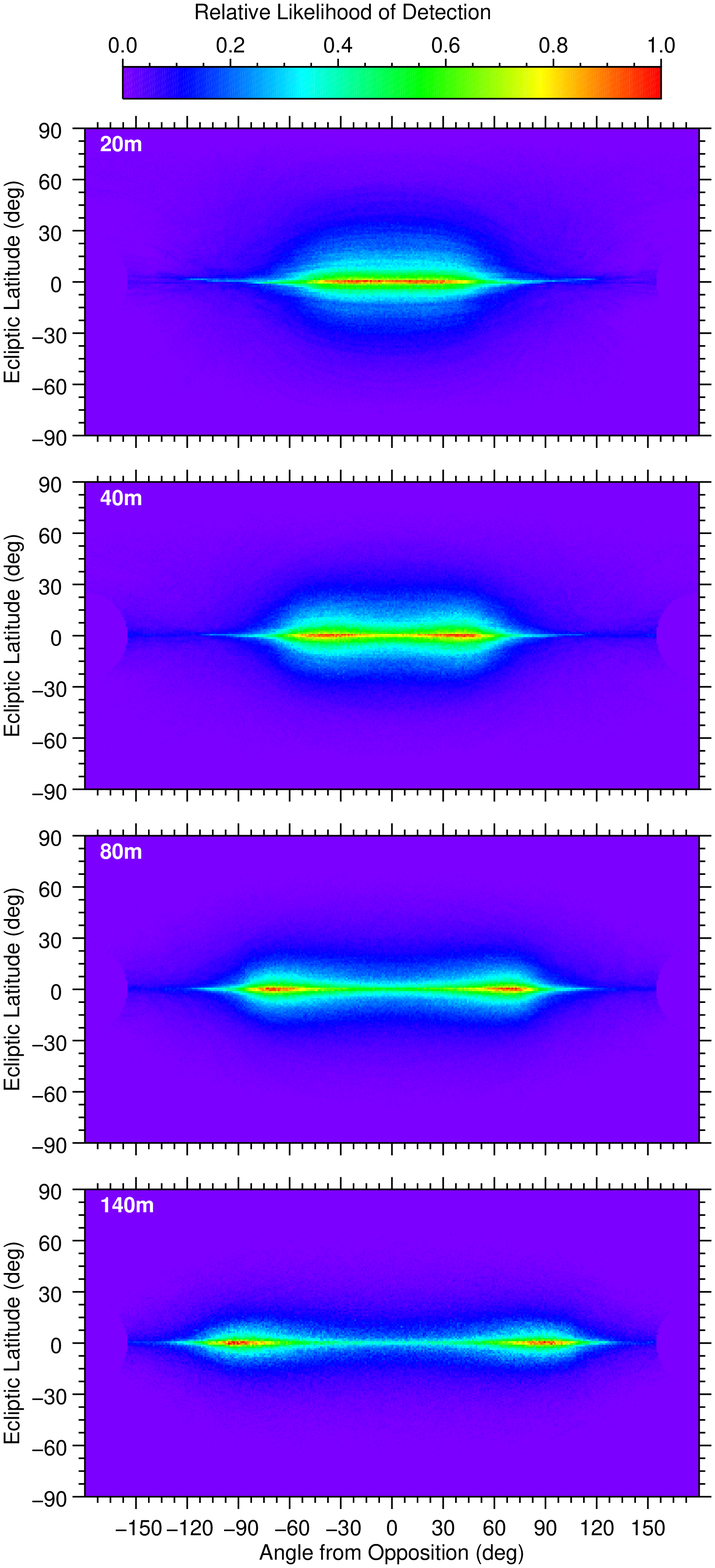}
\figcaption[fig-vimp-v.eps]{\label{fig-vimp-v}\scriptsize
Detectability map for virtual impactors with the nominal Sentinel
design (V) for different target sizes.  Each panel shows where in the sky
a target of the given size can be detected.  The top panel is for the
smallest objects with $D=20$~m and the bottom panel is for $D=140$~m.  Each
map is normalized to 1 at the most likely location.  Points with 25\mydeg\
of the Sun are set to zero.  This sequence clearly shows a transition from
favoring quadrature for detecting larger objects to opposition for detecting
small objects.
}
\end{center}
before with a more limited representation
for the optical \citep{jed96,che04}.  The patterns we show will also
hold for optical observations but with shifts relative to the
\begin{center}
\includegraphics[scale=0.47,trim=80 20 50 30]{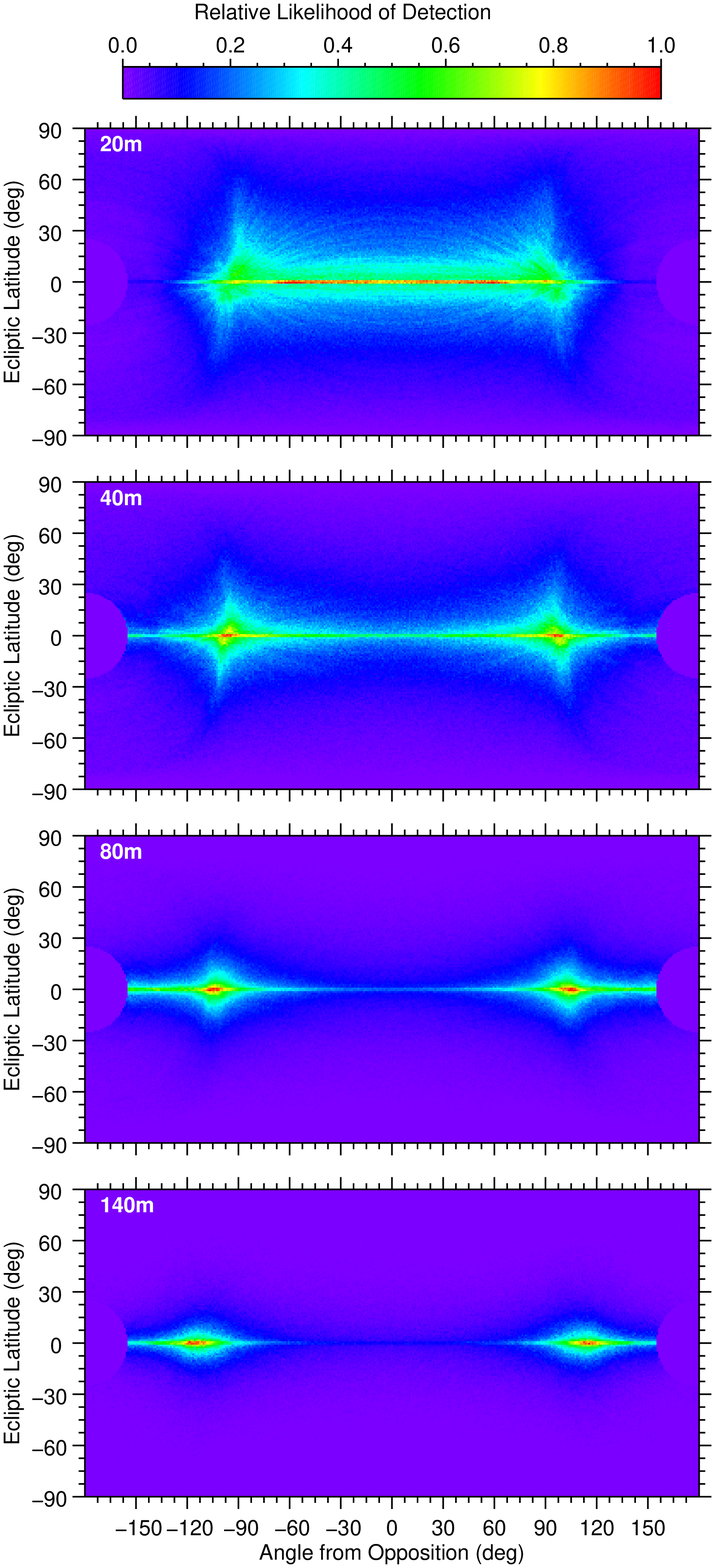}
\figcaption[fig-vimp-e.eps]{\label{fig-vimp-e}\scriptsize
Detectability map for virtual impactors with mission option E
for different target sizes.  Each panel shows where in the sky
a target of the given size can be detected.  The top panel is for the
smallest objects with $D=20$~m and the bottom panel is for $D=140$~m.  Each
map is normalized to 1 at the most likely location.  Points with 25\mydeg\
of the Sun are set to zero.  At this heliocentric distance, detections are
more likely near quadrature with a increasing preference for lower solar
elongation angles as the object size increases.
}
\end{center}
absolute sizes as well as albedo.

Figure \ref{fig-neo-e} shows detection maps for NEAs when surveying
from an Earth-like orbit.  The overall pattern is the same in the two
cases: small objects are preferred at opposition and larger objects are
preferred at lower solar elongations.  There is a significant enhancement
for larger objects as the solar elongations decreases.  Once again,
the best region for detection shifts from low to high solar elongation
as the object size decreases.  The region within 25\mydeg\ is suppressed
in all these figures to suppress this extreme low elongation region to
make it easier to see the structure in the rest of the sky.

Figures \ref{fig-vimp-v} and \ref{fig-vimp-e} show the same detection
maps but based on the virtual impactor population from \citet{che04}
and \citet{ver09}.  As with the NEAs, there is a clear preference
for opposition when searching for small objects and quadrature for
larger objects.  The peak at very low solar elongation is missing in
this population and is due to the much more restrictive
range of orbits and thus heliocentric position.  The virtual impactor
population detection zone also differs from the NEA population in being
more heavily concentrated along the ecliptic.

These figures suggest that one might significantly enhance the survey
detection rate by optimizing the search region.  However, these results
also show that such optimization will not be equally effective at all
sizes.  Note that this visualization only summarizes where objects can
most readily be seen.  The Sentinel survey is designed for self-followup
of its discoveries and performance metrics are built on getting a good
catalog of orbits.  We will return to this important point later in
section \ref{sect-disc}.

\subsection{Modified Field of Regard}

Several options for modifications to the FOR were investigated.
Guided by the zodiacal flux show in Fig.~\ref{fig-zodi}, perhaps the
survey could be more effective by always working closer to opposition and
thus avoid the higher-background regions closer to 90\mydeg\ from the Sun.
Other variations we tried were to avoid the ecliptic poles where the sky
density of objects is lower, FOR's that always keep the telescope
pointed more nearly perpendicular to the spacecraft-Sun line, and
extremely narrow strips near opposition.

Table~\ref{tbl-for} provides some selected results from the FOR variations
examined.  The simulations are run for the baseline mission profile
(eg. Venus-like orbit) with only changes to the FOR\null.  The first two
columns give the range in solar elongation and the ecliptic latitude range
covered.  The last two columns indicate the cycle time in days ($\Delta
t$) and the number of distinct cycles in the 6.5-year survey period
(Nt).  The rest of the table gives the differential completeness ($d$)
of both D=140m and D=30m in percentage units for three different classes
of objects.  Case \#1 shows the performance of the baseline survey.
The performance differences clearly depend on the type and size of object
considered.  When considering either size of NEAs, the peak performance is
achieved by the baseline case.  The same is true for PHAs and impactors.
The results for impactors do show a tie in best performance between
the baseline (\#1) and a FOR more tightly confined to the ecliptic (\#4)
at D=140~m with only a one percentage point increase in completeness
at D=30~m.  The quadrature cases (13--16) are definitely worse than the
baseline and the opposition sliver cases (9--12) are particularly bad.

\begin{deluxetable}{cccccccccccc}
\tablecaption{Completeness versus Field of Regard\label{tbl-for}}
\tablewidth{0pt}
\tablehead{
& & & \multispan2{\hfil NEA\hfil}&
\multispan2{\hfil PHA\hfil}& \multispan2{\hfil Impactors\hfil}&
\colhead{$\Delta$t}& \\
\colhead{Case}&
\colhead{Sel}&
\colhead{Lat}&
\colhead{140}&
\colhead{30}&
\colhead{140}&
\colhead{30}&
\colhead{140}&
\colhead{30}&
\colhead{(days)}&
\colhead{Nt}
}
\startdata
 1&  80--180& $\pm$90& 69& 4.2& 74& 6.4& 88&   19&  28&   85\\
 2&  80--180& $\pm$60& 61& 2.9& 69& 4.4& 86&   15&  24&   98\\
 3&  80--180& $\pm$45& 62& 2.8& 71& 4.9& 87&   16&  20&  120\\
 4&  80--180& $\pm$30& 61& 2.8& 73& 6.0& 88&   18&  14&  170\\
 5&  80--180& $\pm$20& 55& 2.2& 70& 5.1& 87&   17&  10&  248\\
 6&  80--180& $\pm$10& 43& 1.4& 62& 3.8& 81&   13&   5&  488\\
 7& 130--180& $\pm$90& 46& 2.2& 55& 4.4& 80&   15&  14&  170\\
 8& 150--180& $\pm$90& 25& 0.8& 34& 1.9& 65& 09.3&   8&  283\\
 9& 160--180& $\pm$90& 16& 0.3& 25& 1.0& 55& 06.5&   6&  424\\
10& 170--180& $\pm$90&  7& 0.1& 13& 0.5& 37& 03.5&   3&  848\\
11& 170--180& $\pm$45&  8& 0.1& 14& 0.5& 38& 03.8&   2& 1199\\
12& 170--180& $\pm$30&  8& 0.1& 14& 0.5& 39& 04.0& 1.4& 1695\\
13&  70--140& $\pm$90& 65& 3.4& 71& 4.0& 85&   12&  20&  122\\
14&  70--140& $\pm$45& 62& 2.2& 70& 3.2& 85&   12&  14&  172\\
15&  45--120& $\pm$90& 58& 2.7& 65& 2.6& 78& 07.4&  21&  113\\
16&  45--120& $\pm$45& 59& 1.9& 69& 2.6& 83& 07.5&  15&  160\\
\enddata
\tablecomments{\scriptsize
The FOR for each case is defined by the range of the solar elongation (Sel)
and ecliptic latitude (Lat) covered on the sky.   The completion values
shown for three sample populations and two sizes (in meters) are given 
as a percentage of that population.  The time required to cover the
FOR is given under $\Delta t$ and Nt is the number of times the FOR is
covered during the 6.5 year survey duration.
}
\end{deluxetable}

%IDL> @for_table

The information presented in Fig.~\ref{fig-neo-v}--\ref{fig-vimp-e}
compared to the results in Table~\ref{tbl-for} brings up a very
important point.  Sky-plane detection maps are very useful to show where
a single-epoch observation is most effective at detecting objects.  These
have been used very successfully to guide ground-based survey efforts.
The detection maps are sufficient to guide the observation planning in
the case where the first detection is the sole output of a survey and
followup observations are separately handled.  When a survey does its own
followup the observing constraints are different.  For example, if an
object had such an orbit that it could only be seen once in the search
area, the system would not get a followup observation and the object
would not be considered cataloged with a good enough orbit.  The survey
simulation combines the detection map with the geometric realities
of the followup observations needed.  As seen in Table~\ref{tbl-for},
the survey performance optimization based on FOR is much weaker than
expected from the detection maps alone.

\subsection{Exposure Time Modifications}

There are two consequences of changing the exposure time: changed time
between visits and different sensitivity levels.  Longer exposures go
deeper except when limited by trailing losses.  This change will be
more effective at quadrature where the target motion is lower than at
opposition.  After extensive testing with the simulator, the benefit
of changing exposure time is minimal.  It appears to be most useful as
a trade-off against the FOR and the implied cycle interval.  The baseline
exposure time of 180 seconds when coupled with the choice of telescope
and detector characteristics is an excellent choice and we find little
reason to modify this aspect of the survey.

\section{Simulation Results}

This section contains select results from the survey simulations.  In the
figures and discussion to follow the label `V' is used for the baseline
Sentinel mission for its Venus-like orbit, `VE' is used for the same
mission profile but with an orbit with perihelion at Venus and aphelion at
Earth, `E' is again the Sentinel mission hardware at
the Earth's orbit, and `N' is for an adaptation of the NEOCam mission
concept as described in \citet{mai15a} where we use their FOR design
and observatory orbit with the rest of the Sentinel mission parameters.
In the case of all Sentinel concepts, an object must be observed with a
minimum of 28 days orbital arc to be considered a cataloged object and
count towards completeness.  However, the native cadence of NEOCam is
20 days for an similar criterion.  During our modeling of the NEOCam
FOR we also used this shorter value.  This accommodation gives a slight
advantage to the NEOCam FOR scenario with a correspondingly weaker orbit
determination for a minimum observational set on an object.

\subsection{Baseline Sentinel Mission}

Figure~\ref{fig-basecumc} provides a summary view of the survey
performance of the baseline Sentinel mission.  In this graph, and the
survey summaries to follow, the cumulative detection performance is shown
against target size.  The performance here is modeled by physical size
and the $H$ magnitude is derived for the purpose of the plot using a
standard conversion of $H_V$=22.1, geometric albedo=14\%, and $D=140$~m
\citep{pra12}.  The different colored lines show a wide array of objects
with different dynamical groupings.  The basic NEA population from
\citet{bot02} is drawn with a black line.  All other types are varying
subsets of the NEA population as described in section~\ref{sect-samp}.
Some general trends are evident in these results.  The Amor population
performance is the worst of all at small sizes though it is quite good at
larger sizes.  This curve shows the transition from a geometry-limited
survey at large sizes and a sensitivity limited survey at small sizes.
In contrast, the Aten population is systematically better observed at all
sizes due to being closer to the observatory and thus easier to observe.
The Atira population is even better observed at small sizes but there is
a break in the completeness curve around $D>60m$ where the completeness
tops out at 80\%.  This plateau is a consequence of those objects that
happen to remain unobservable for the entire mission due to geometry,
that is, Sentinel and those objects are always on the opposite sides
of the Sun from each other during the 6.5-year survey.  A longer survey
would top out at a higher level.  Of these populations, the baseline
survey is most effect on the ARRM target sample and reaches 100\%
completeness around $D>65$~m.
\begin{center}
\includegraphics[scale=0.29,trim=80 30 50 30]{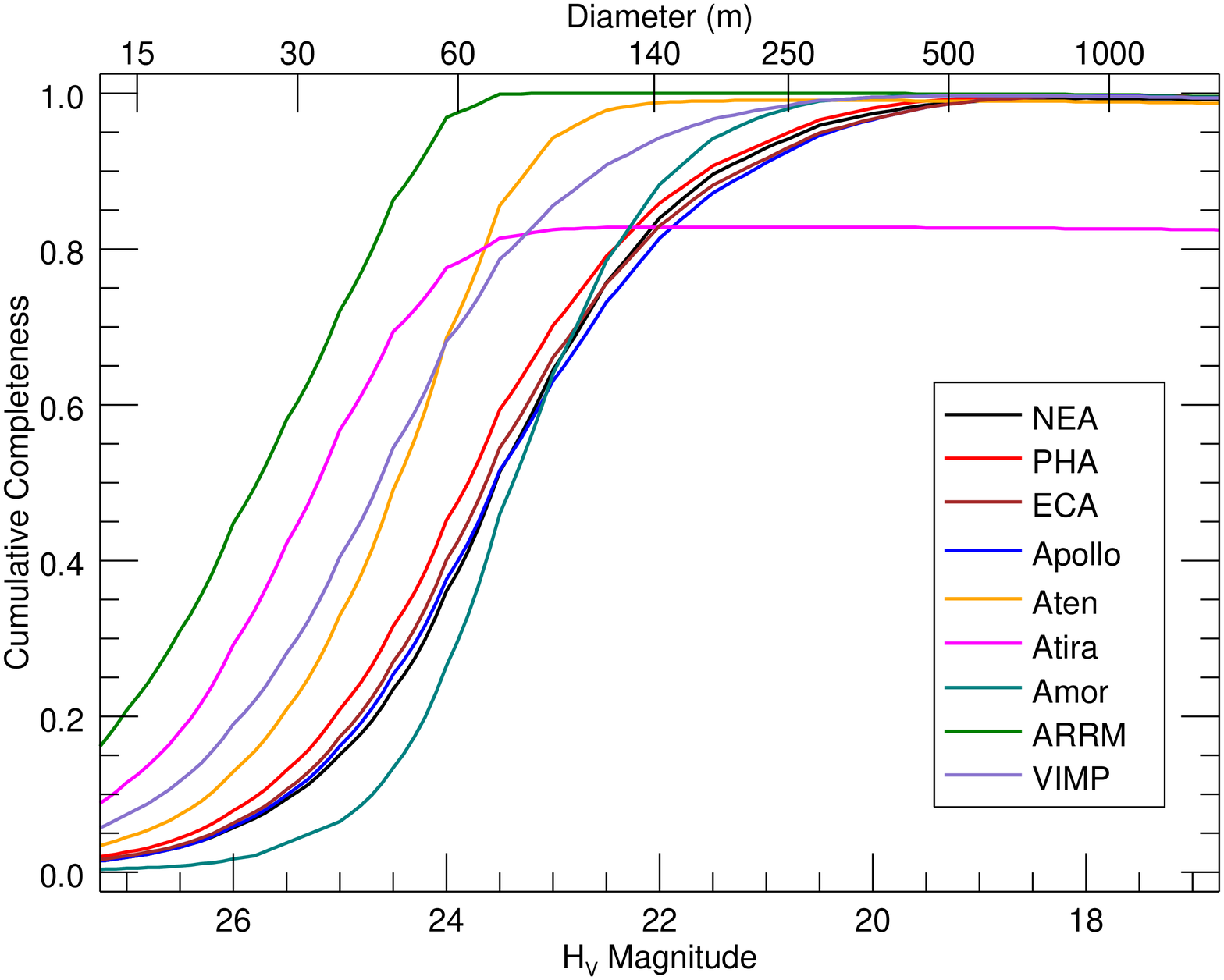}
\figcaption[figbasecumc.eps]{\label{fig-basecumc}\scriptsize
Cumulative survey results for the baseline Sentinel mission.
The mission duration is 6.5 years and the different curves show
different sub-populations of the NEA orbit distribution.
$H_V$ is computed from the input diameter using a 14\% geometric albedo.}
\end{center}

Another instructive view of the survey performance is to look at the
cumulative completeness as a function of time.  Figure~\ref{fig-time}
shows such a curve for NEAs (dashed lines) and the Virtual Impactor (VIMP)
sample (solid lines).  Completeness curves are shown for four different
limiting sizes.  For the D$>140$~m impactor population, 90\% completeness
is reached in about 4 years.  If the target sample
\begin{center}
\includegraphics[scale=0.29,trim=80 30 50 30]{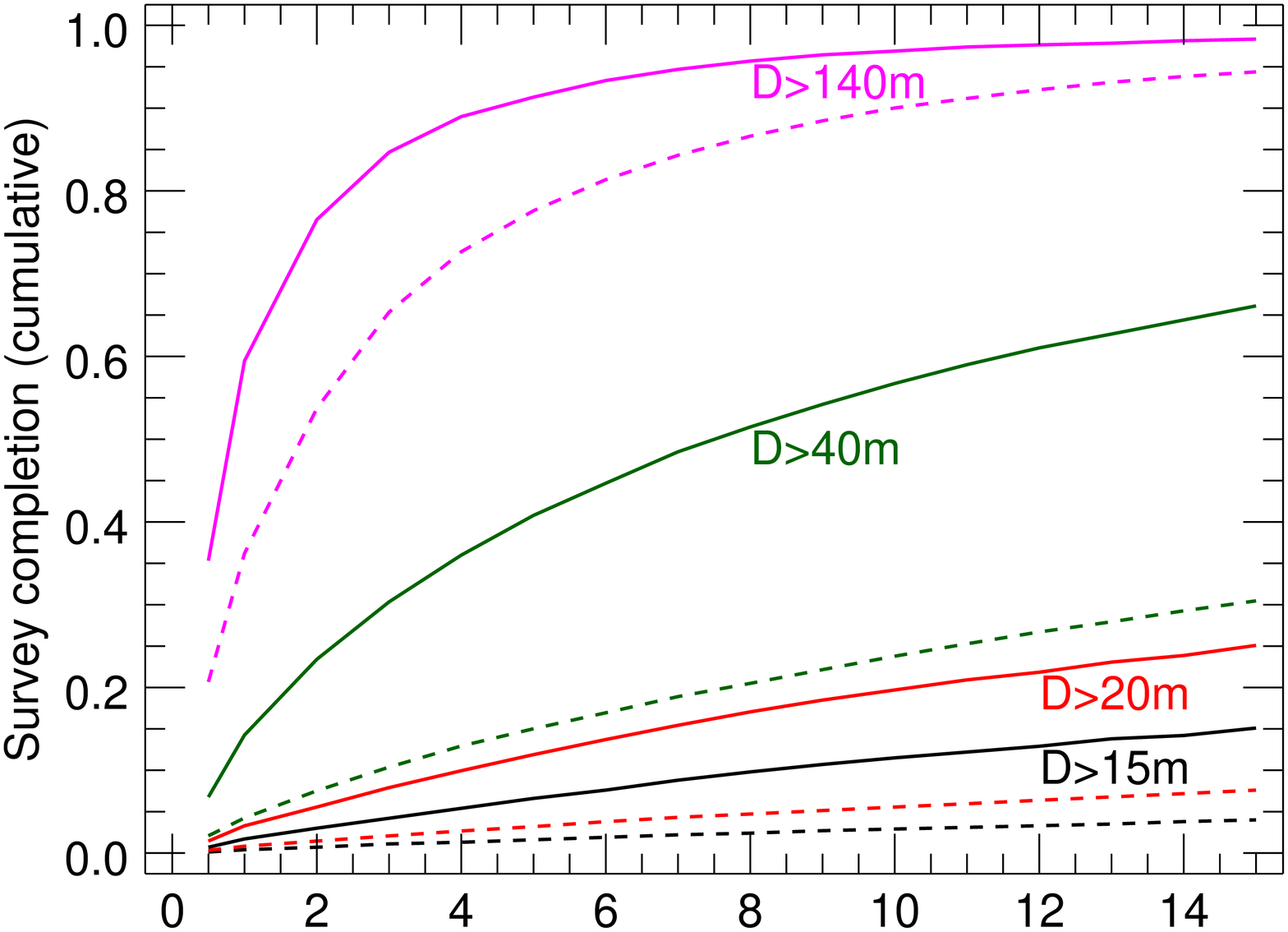}
\figcaption[figtime.eps]{\label{fig-time}\scriptsize
Cumulative survey results for the baseline Sentinel mission as a function
of survey duration for a variety of limiting sizes.  The dashed curves
show the results for the full NEA population.  The solid curves show
the results based on a sample of virtual impactors.}
\end{center}
is the entire NEA population it takes nearly 10 years
to reach this same completeness level.
This plot clearly shows an asymptotic tail for the largest objects that
is due to orbital geometry beyond the control of the survey (eg., long
orbital periods for some objects).  The overall completeness level is
clearly seen to decrease when considering smaller sized objects.  The
cases shown for $D>$15-20~m objects show these to be in a roughly linear
regime where surveying for twice as long will return twice as many
discoveries.  In this case, two survey instruments with uncorrelated
vantage points will detect new objects twice as fast.

\begin{center}
\includegraphics[scale=0.29,trim=80 30 50 30]{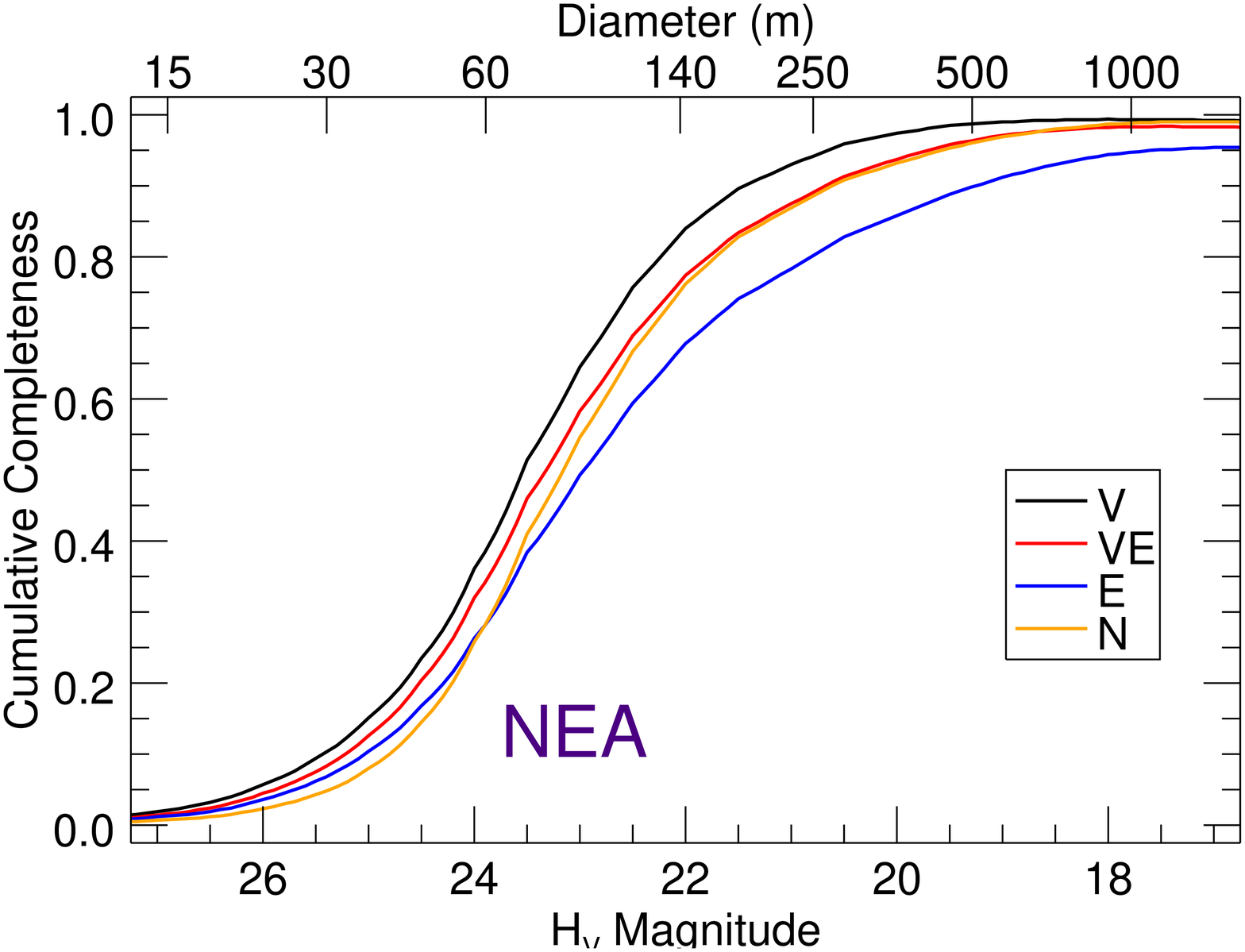}
\figcaption[figneo.eps]{\label{fig-neo}\scriptsize
Cumulative survey results for NEAs from a 6.5 year survey.
$H_V$ is computed from the input diameter using a 14\% geometric albedo.}
\end{center}

It is instructive to show sub-population results individually
for a variety of mission profiles.  The first of these is shown in
Fig.~\ref{fig-neo} for the \citet{bot02} NEA population.  We use ``V'' to
denote the baseline survey which has a Venus-like orbit.  In this case,
the baseline mission has
the best performance at all sizes for these
four mission options but the variation in survey performance is small.
The worst concept of this set is for mission option E which reaches a
completeness of $\sim$65\% for $D>140$~m.  It is clear that this vantage
point requires a significant change in mission design, such as the NEOCam
FOR concept, to get better performance.  For sizes below 60m, the NEOCam
FOR survey has the lowest performance of these four options.  Clearly,
if the point of a survey is a measurement of the general population,
all of these surveys will constrain the population down to $D=20$~m
after careful de-biasing.  If planetary defense is a goal where the
total number detected is important, there are discriminating choices to
be made from these results.

\begin{center}
\includegraphics[scale=0.29,trim=80 30 50 30]{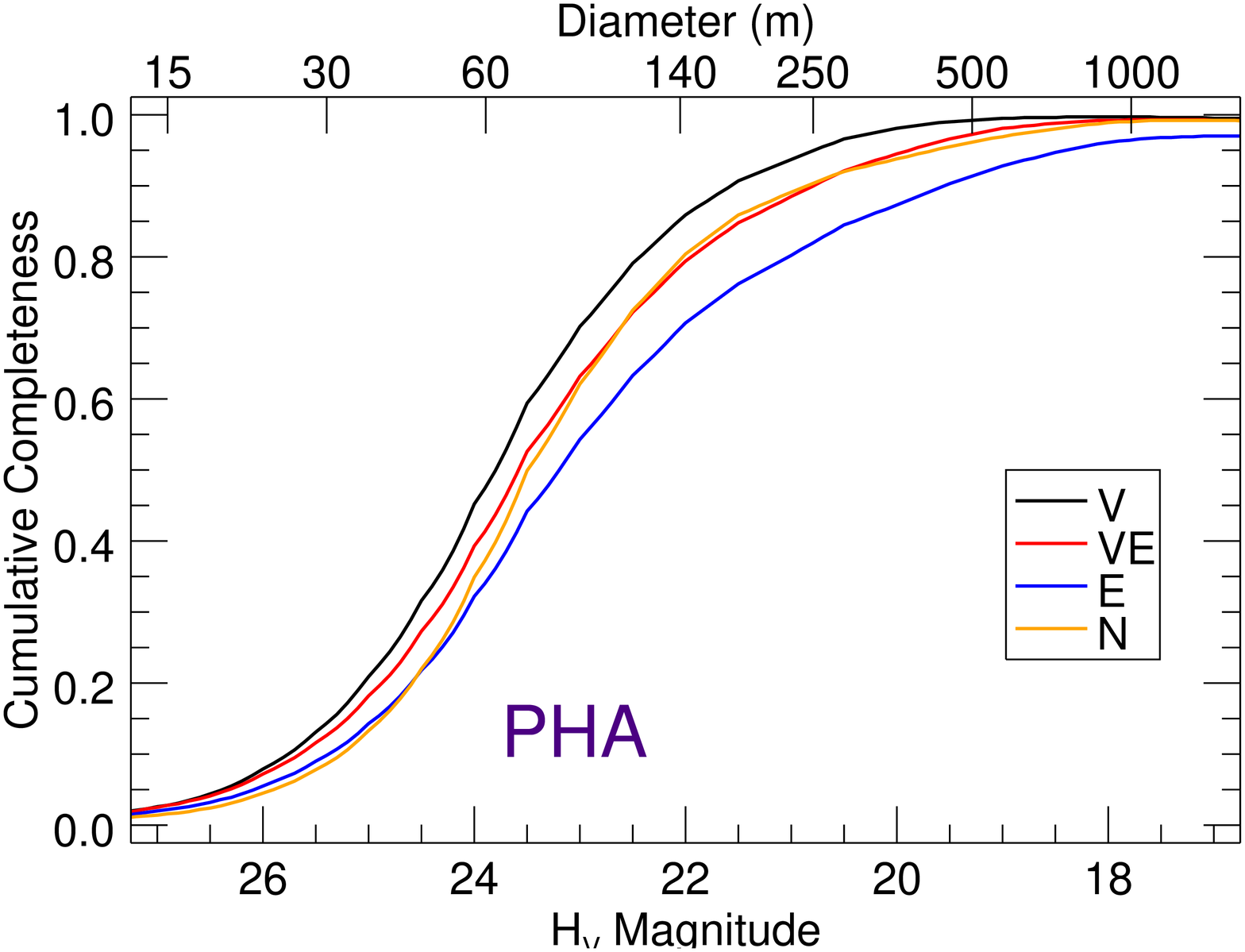}
\figcaption[figpho.eps]{\label{fig-pho}\scriptsize
Cumulative survey results for PHAs from a 6.5 year survey.
$H_V$ is computed from the input diameter using a 14\% geometric albedo.}
\end{center}

Figure~\ref{fig-pho} shows the survey performance for potentially
hazardous objects defined by an Earth MOID of 0.05 AU or less.  As with
the NEA population, these curves are all more similar than they are
different.  Nonetheless, the spread of performance between Sentinel in
a Venus-like orbit and option E is slightly larger here for $D=50$~m
objects.  Once again, there is a drop in the NEOCam FOR performance
below 60m relative to the other options.

The least relevant sub-population for near-term threats to the Earth are
the Amors.  Again, these curves really are more similar than different
but the consequence of this orbit sample is evident in survey performance
(see Fig.~\ref{fig-amo}).  The baseline Sentinel has the best performance
for $D>120$~m but has the lowest performance below 50~m.  The NEOCam
FOR performance is on the lower
\begin{center}
\includegraphics[scale=0.29,trim=80 30 50 30]{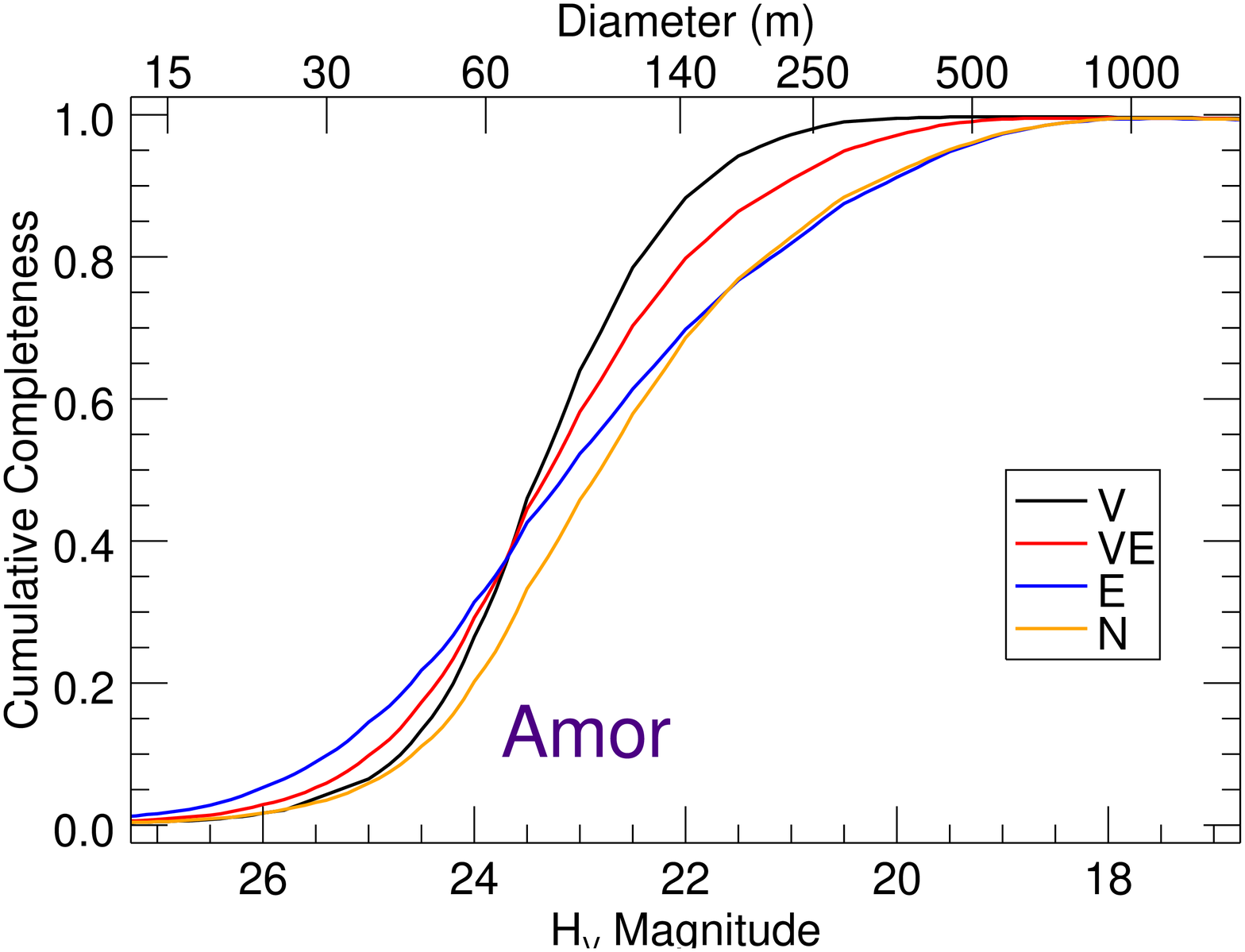}
\figcaption[figamo.eps]{\label{fig-amo}\scriptsize
Cumulative survey results for Amors from a 6.5 year survey.
$H_V$ is computed from the input diameter using a 14\% geometric albedo.}
\end{center}
boundary of all scenarios and reveals
an interesting difference from
large object performance where
it is
identical to option E and small object performance where it is identical
to the baseline Sentinel.  We conclude that at sizes of $D>160$~m, the
survey performance is insensitive to the FOR or the choice of orbit for
the mission.  The baseline Sentinel FOR works very well in this case
at an Earth vantage point because it looks directly outward where this
population is most likely to be found.  By being closer, it is more
effective on small objects.

\begin{center}
\includegraphics[scale=0.29,trim=80 30 50 30]{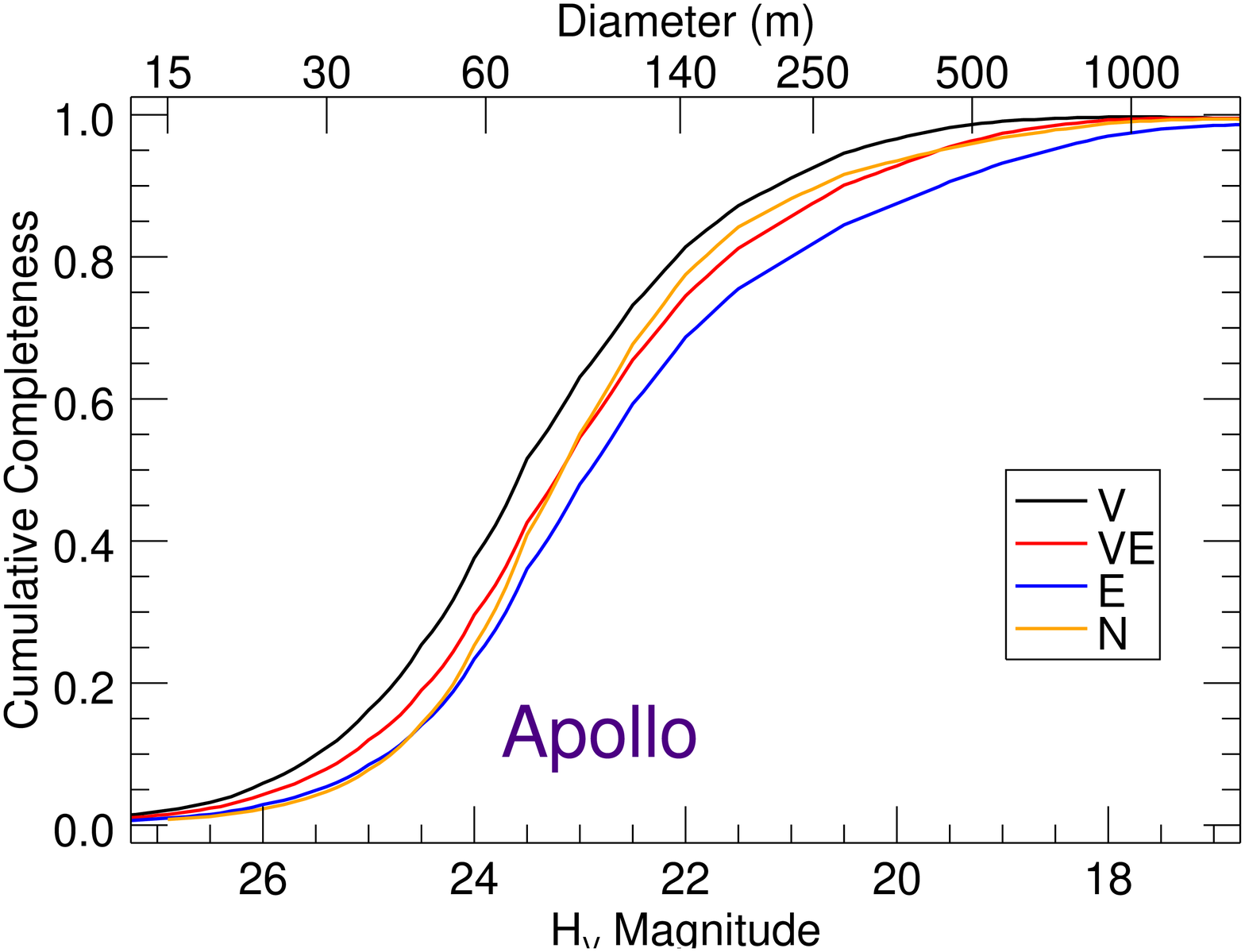}
\figcaption[figapo.eps]{\label{fig-apo}\scriptsize
Cumulative survey results for Apollos from a 6.5 year survey.
$H_V$ is computed from the input diameter using a 14\% geometric albedo.}
\end{center}

There is very little different about the survey performance for Apollos
(see Fig.~\ref{fig-apo}).  Once again, Sentinel at Venus and option
E bound the range of survey performances with the baseline mission having
slightly better performance.

\begin{center}
\includegraphics[scale=0.29,trim=80 30 50 30]{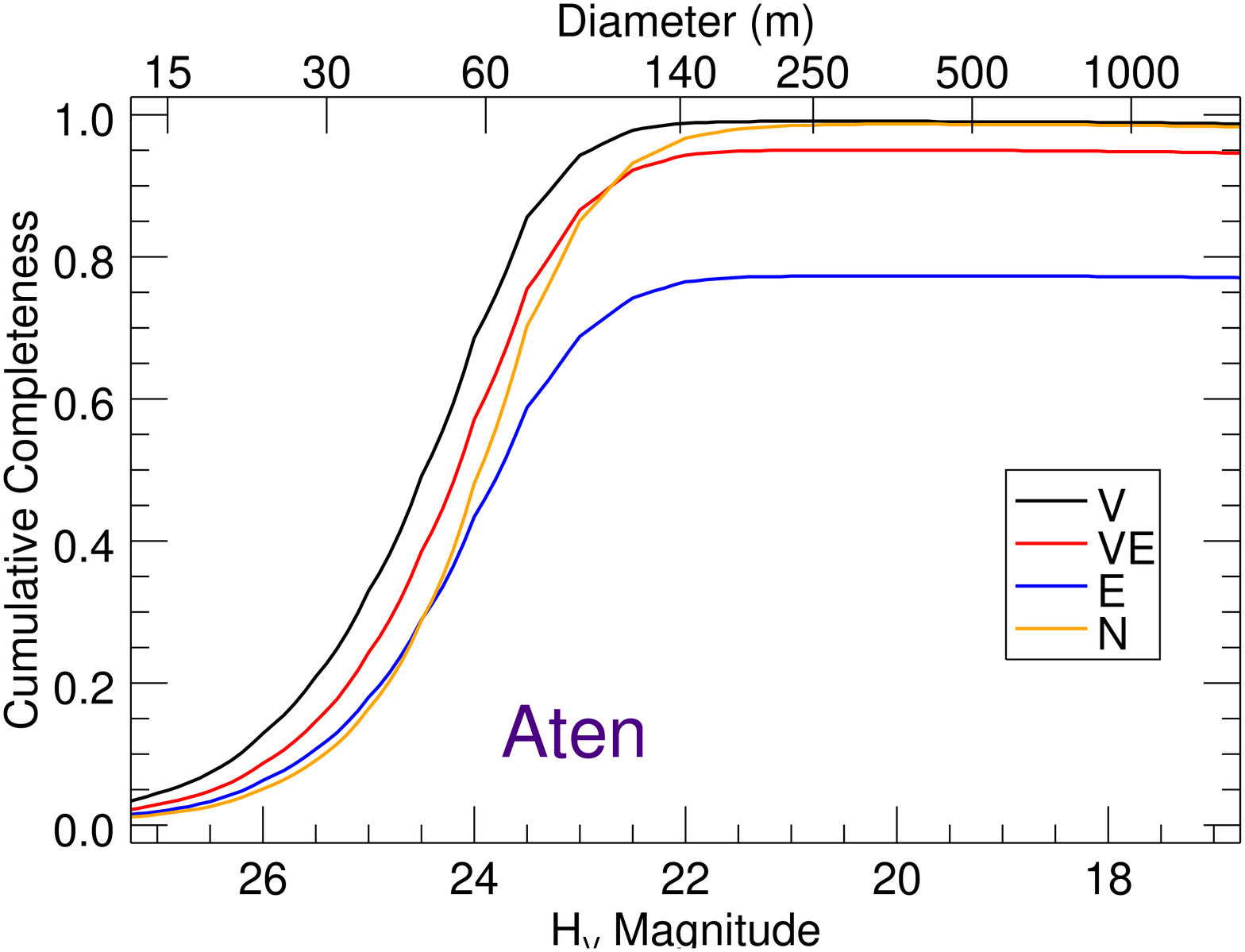}
\figcaption[figate.eps]{\label{fig-ate}\scriptsize
Cumulative survey results for Atens from a 6.5 year survey.
$H_V$ is computed from the input diameter using a 14\% geometric albedo.}
\end{center}

Moving inwards, the performance for Atens is shown in Fig.~\ref{fig-ate}.
Both the baseline Sentinel FOR and NEOCam FOR surveys work extremely well
for $D>140$~m with essentially 100\% completeness.  Fundamental geometry
limitations seriously affect mission option E\null.  The missed
objects just never cross into the FOR\null.  These objects are
viewable only at small solar elongations where a FOR like NEOCam provides
good coverage.  For these large sizes, NEOCam is a very
effective option but this advantage drops off to smaller sizes.

\begin{center}
\includegraphics[scale=0.29,trim=80 30 50 30]{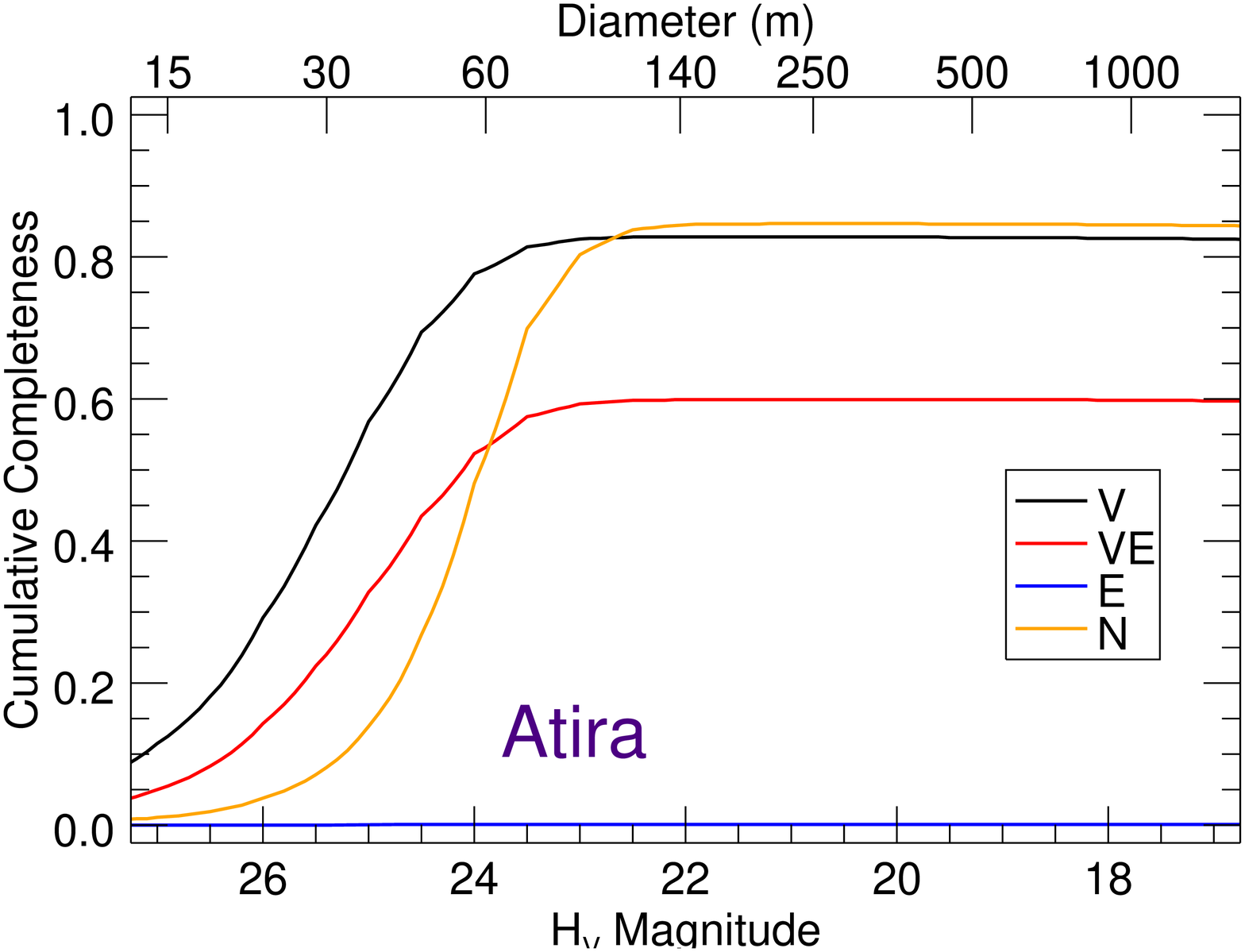}
\figcaption[figati.eps]{\label{fig-ati}\scriptsize
Cumulative survey results for Atiras from a 6.5 year survey.
$H_V$ is computed from the input diameter using a 14\% geometric albedo.}
\end{center}

\vskip-10pt
Completing the suite of NEA sub-groups are the Atiras, shown in
Fig.~\ref{fig-ati}.  These are the objects orbiting closest to the
sun and all surveys have geometric limitations for studying these objects.
The worst is mission option E which never sees this population
at all and the VE option is lower as well because time is spent
near the Earth where the Sentinel FOR is less effective.
Just as with Apollos, the baseline Sentinel FOR and NEOCam FOR perform
well and are effective for large objects and the NEOCam FOR performance drops
off more quickly for small objects.

\begin{center}
\includegraphics[scale=0.29,trim=80 30 50 30]{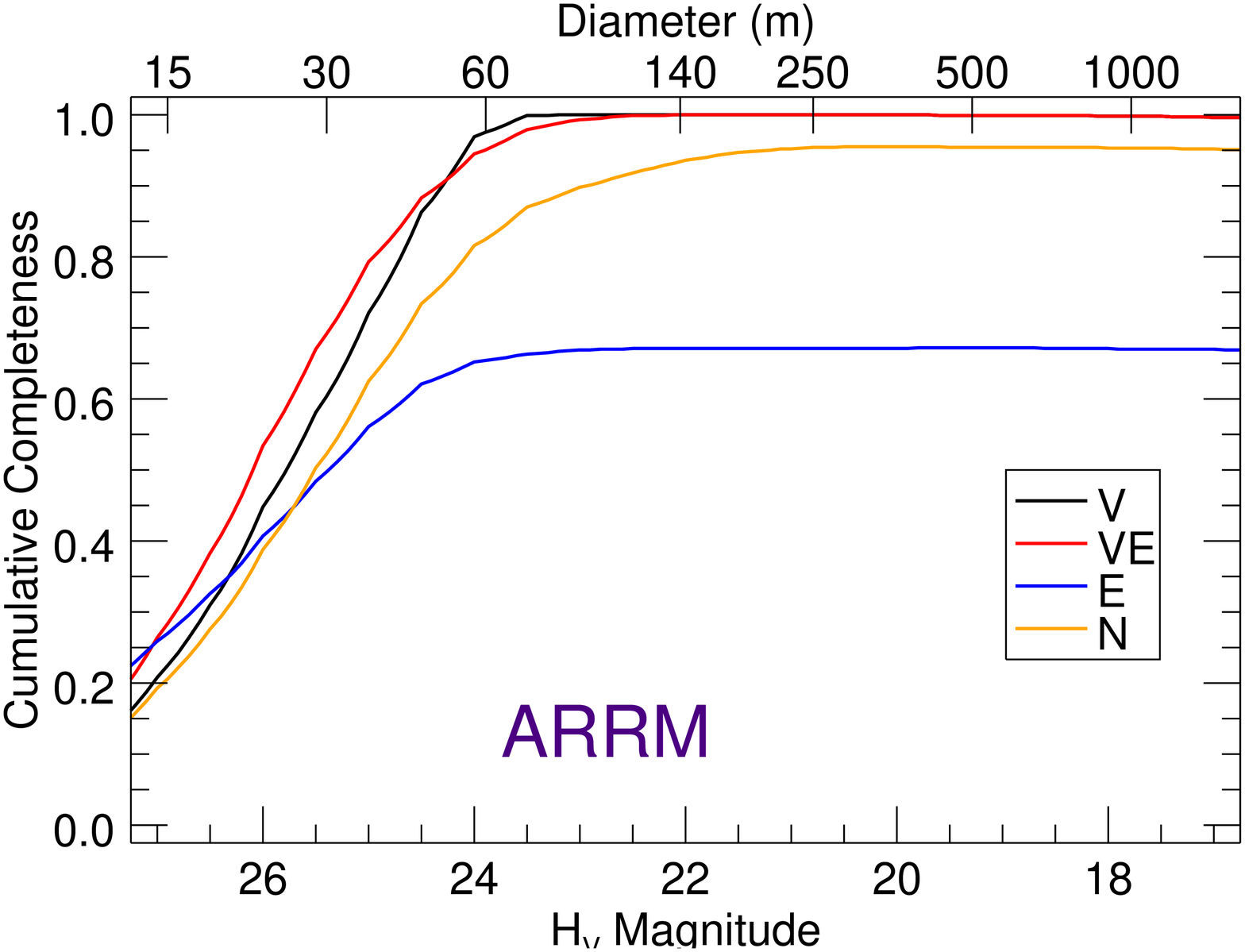}
\figcaption[figarrm.eps]{\label{fig-arrm}\scriptsize
Cumulative survey results for ARRM targets from a 6.5 year survey.
$H_V$ is computed from the input diameter using a 14\% geometric albedo.}
\end{center}

A special sub-class of objects, labeled as ARRM targets, is shown in
Fig.~\ref{fig-arrm}.  These objects are special in that they provide low
launch energy rendezvous orbits for a spacecraft originating at the Earth.
In this case, all options find a significant fraction of these objects.
Two of the options, Sentinel-V and Sentinel-VE, all are more than 95\%
complete for $D>60$~m and a significant fraction are seen by all mission
options down to 15m sizes.  We see once again the clear benefit of the
NEOCam FOR at an Earth orbiting location compared to option E, which is
blind to about 30\% of this group.

The final category of virtual impactors is shown in Fig.~\ref{fig-vimp}
and is the sample of orbits from \citet{che04} and \citet{ver09}.
Once again we see that all mission options are similarly effective at
large sizes.  The baseline Sentinel option has the best performance at
all sizes and is almost twice as effective as NEOCam at $D>25$~m.

\begin{center}
\includegraphics[scale=0.29,trim=80 30 50 30]{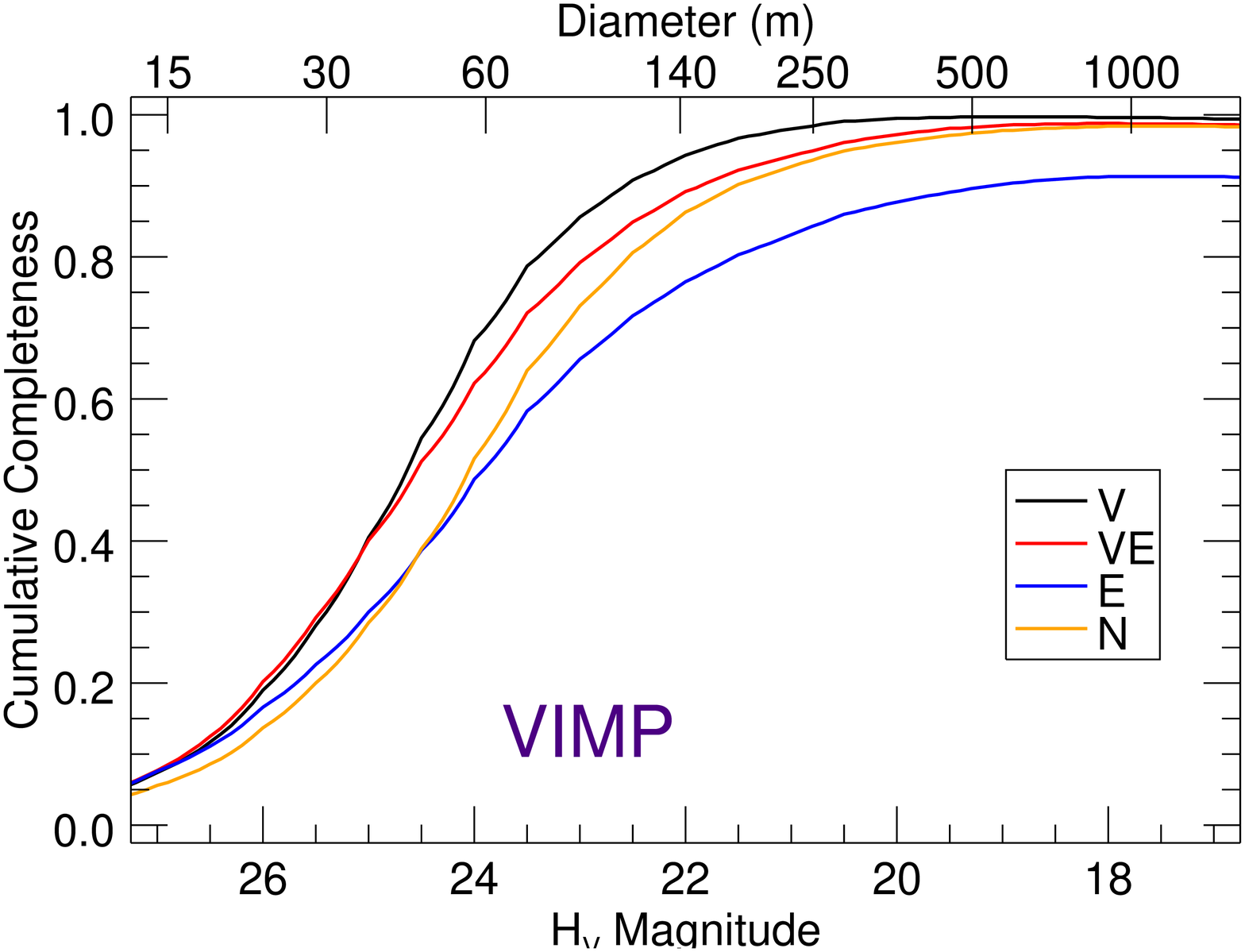}
\figcaption[figvimp.eps]{\label{fig-vimp}\scriptsize
Cumulative survey results for Virtual Impactor targets from a 6.5 year survey.
$H_V$ is computed from the input diameter using a 14\% geometric albedo.}
\end{center}

All of the preceding performance plots show completeness alone and
are very instructive to see the effectiveness of a given strategy.
However, the total number to be detected is equally important.  To get
a more complete picture we combine the normalization factors from
Table~\ref{tbl-norm} with the completeness curves.

From a planetary defense perspective, the most important of these
performance curves is the actual impactor population.  By combining
the results of Fig.~\ref{fig-vimp} and the normalization factor we
can show an estimate of the absolute number of objects to be found.
Figure~\ref{fig-impact} shows the result of this calculation.  Here we
plot the cumulative number of objects that will hit the Earth in the next
100 years as a function of the limiting size.  The solid line is the
number of objects that are there to be found.  Note that this graphic
has zoomed in exclusively on the small end of the size distribution.
The total number reaches a value of one at $D>30$~m.  For larger size
cut-offs the expected number of actual impactors to be found is much
less than one.  Clearly, the most likely to hit Earth in the next 100
years is in this sub-30~m size range.  This is a direct consequence of
the overall low impact probability that must be overcome by reaching
a point in the size distribution where the number of objects is large
enough to compensate.  There are very few objects out there to be found
that will hit in the next 100 years.  These calculations indicate that
the total number is $\sim$7 for $D>15$~m.  For $D>20$~m this number has
dropped to 3\null.  These calculations imply an impact every 30 years
on average for objects down to 20~m in size, roughly consistent with the
observed impact rate \citep{bro13}.  Also shown on Fig.~\ref{fig-impact}
are estimates (left axis label) for the number of impactors detected
by the baseline Sentinel mission (black dashed curve) and for the VX
option (magenta dot-dash curve).  The VX option is an improvement of
about a factor of two in performance but likely much more expensive
than a factor of two due to increased aperture and mission lifetime.
With this example in mind it might be more cost-effective to have two
Sentinel telescope rather than scale up a single observatory.  At the
low end of the size range the completeness is low and two observatories
will catalog twice as many objects.  The scale on the right provides an
estimate of how many objects pass within 10~R$_{\earth}$ in 100 years.
The choice of 10~R$_{\earth}$ has no special meaning other than such
objects still get really close to the Earth and would most likely generate
attention without hitting while being a simple number.   This example
illustrates that there will
\begin{center}
\includegraphics[scale=0.29,trim=80 30 50 30]{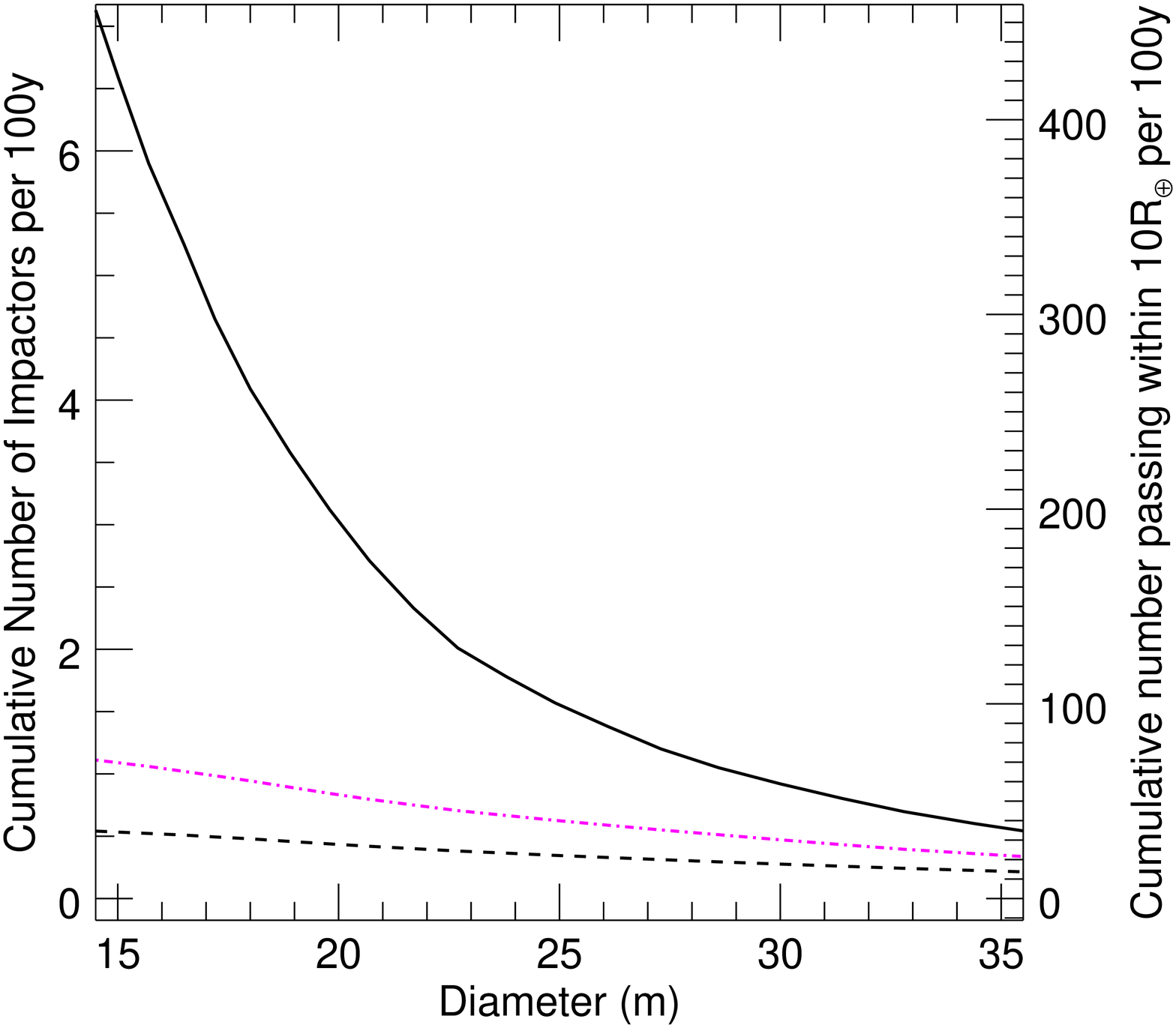}
\figcaption[figimpact.eps]{\label{fig-impact}\scriptsize
Cumulative survey results for Earth impactors from the baseline Sentinel
mission.  This calculation is for a 6.5 year survey.  The solid line
is the cumulative number of impactors in 100 years.  The black dashed curve
is the estimated number of these impactors that would be detected for the
baseline Sentinel mission.   The magenta dot-dash curve is for the enhanced
(VX) Sentinel survey.
The scale on the right shows the estimate of how many objects
come with 10~R$_{\earth}$ over the same time.
$H_V$ is computed from the input diameter using a 14\% geometric albedo.}
\end{center}
be quite a few objects whose orbits will
require close scrutiny to ensure that they are on non-impacting orbits.
This estimate counts each object once and does not track multiple passes
by the same object.  A summary of key results in this figure is provided
in Table~\ref{tbl-impact} along with another common case of objects
passing within a lunar distance from Earth.  The columns labeled ``N''
indicate the total cumulative number of objects for that size limit based
on the size distribution.  The columns labeled $N_{det}$ indicate the
number that would be detected by each survey case.  This table clearly
shows the increasing completeness in the survey with increasing size while
also illustrating the rapid decline in the total number that either hit
or get close to the Earth over a 100-year period.

\begin{deluxetable}{cccccccccl}
\tablecaption{Impactors and Close Flyby Objects in a 100 year time interval\label{tbl-impact}}
\tablewidth{0pt}
\tablehead{
\colhead{Objects}&
\multispan2{\hfil D$>$15m\hfil}&
\multispan2{\hfil D$>$30m\hfil}&
\multispan2{\hfil D$>$50m\hfil}&
\multispan2{\hfil D$>$140m\hfil}&
Case\\
& N& $N_{det}$& N& $N_{det}$& N& $N_{det}$& N& $N_{det}$&
}
\startdata
Impactors& 6.6& 0.53& 0.92& 0.28& 0.18& 0.11& 0.014& 0.014& Baseline\\
Impactors& 6.6&  1.1& 0.92& 0.47& 0.18& 0.14& 0.014& 0.014& VX\\
within 10~R$_{\earth}$&
           420&   34&   59&   18&   12&  7.0&  0.89&  0.89& Baseline\\
$<$Lunar distance&
        15,000& 1,200& 2,100& 640& 420&  250&    32&    32& Baseline\\
\enddata
\tablecomments{\scriptsize
This table contains estimates of the absolute numbers of objects in a
category and detected in the baseline and VX surveys.  The estimates
are cumulative numbers above different size cutoffs.  The columns labeled
``N'' is the total number in the category and ``$N_{det}$'' is the number
predicted to be cataloged in a 6.5 year survey.
}
\end{deluxetable}

% values taken from:
% v_vimp_06.5_5.0_050_0.14_cum.cum   baseline
% v_vimp_10.0_5.0_065_0.14_cum.cum   VX

% 10RE/impact  = 63.6
% Lunar/impact = 2270

\begin{center}
\includegraphics[scale=0.29,trim=80 30 50 30]{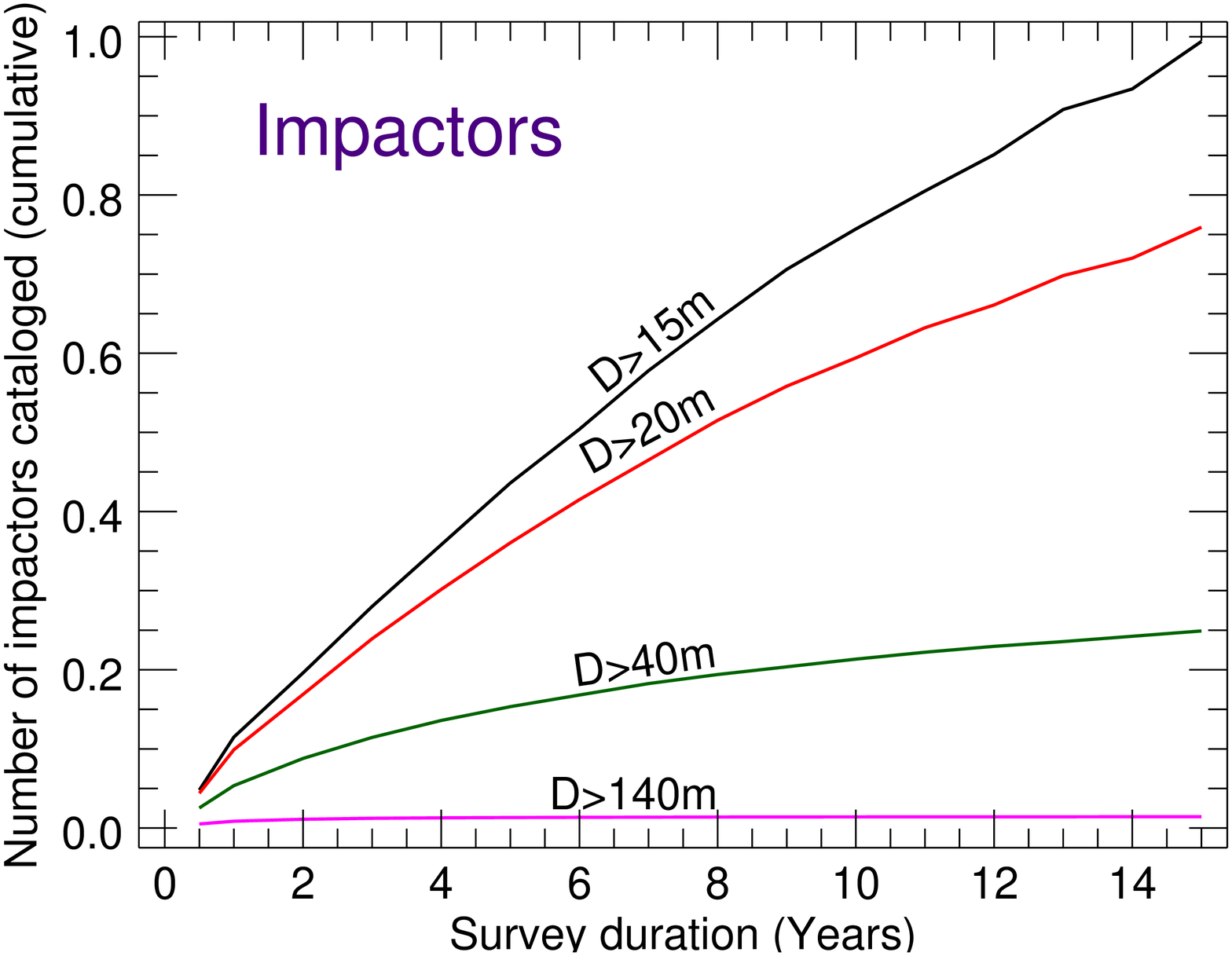}
\figcaption[figvimptimenum.eps]{\label{fig-vimptimenum}\scriptsize
Cumulative survey results for Earth impactors from the baseline Sentinel
mission as a function of survey duration.  Shown here are four curves for
different limiting size cutoff for the survey.  All results show an estimate
of the
cumulative number of objects that would be detected during the survey
that will be an impactor during the next 100 years.  $H_V$ is computed
from the input diameter using a 14\% geometric albedo.}
\end{center}

The expected survey results for impactors is also shown in
Fig.~\ref{fig-vimptimenum}.  In this figure, the number of objects
detected is shown as a function of time for four different limiting sizes.
The expected number of large impactors is very low, predominantly due to
consequences of the size distribution.  Sentinel is very good at finding
large objects ($D>140$~m) but there just aren't that many out there that
are impactors.  This figure clearly shows the need to effectively search
to much smaller sizes to have a significant survey yield of impactors.
Within the scope of this study, Sentinel does as good of a job finding
these objects as you can do with a single observatory.  There are
minor performance improvements possible but these come at considerable
expense, such as working with a bigger telescope.  A more practical way
to significantly improve on impactor discoveries is to have more than one
survey facility and to have them all in different locations so that their
discovery volumes don't overlap.  At the smallest sizes, the completion
fraction is low enough that running two surveys will net twice as many
impactors.  We will discuss this scenario further in the next section.

\section{Simultaneous Operation with LSST}\label{sect-lsst}

Ground-based discovery of NEAs will continue to improve, and the
performance of a space survey must be considered in coordination with
ground-based results. We have extended our space mission IR model
to include ground-based visible observations to permit analysis of
the performance improvement made possible by the addition of a space
survey. We use the performance of the Large Synoptic Space Telescope (LSST)
as a proxy for the performance of all ground-based surveys due to its
high potential for discovery of most of the objects that are within
reach of ground-based telescopes. We have used this model to evaluate
Sentinel's incremental discovery capabilities for several NEA orbital
sub-populations and for various limiting sizes.

\subsection{Performance Model for the LSST}

The design and anticipated performance for the LSST have been published
by the LSST consortium \citep{ive06}. Key parameters that we used to define
its performance are summarized in Table~\ref{tbl-lsst}.

\begin{deluxetable}{cc}
\tablecaption{LSST parameters used for the simulations\label{tbl-lsst}}
\tablewidth{0pt}
\tablehead{
\colhead{Parameter} &
\colhead{Value}
}
\startdata
Detector passband&                                  552-691 nm ({\it r} band)\cr
Detector quantum efficiency&                         0.85\cr
Optical throughput from secondary and mirror losses& 0.40\cr
Detector read noise$^{\dagger}$&                             25 e$^-$\cr
Telescope diameter&                                  8.4 m\cr
Minimum SNR for detection&                           5\cr
Detector fill factor&                                90\%\cr
\enddata
\tablecomments{\scriptsize
$\dagger$The estimated read noise is 12 electrons per 15-s exposure.  We
multiplied by $\sqrt 2$ to account for two exposures per 30-s integration,
and further multiplied by $\sqrt{9/4}=1.5$ because the oversampling of the
LSST focal plane would lead to a 3x3 pixel detection kernel.
}
\end{deluxetable}

NEA albedo has a strong influence on detectability in the visible
wavelengths.  Based on observations from WISE \citep{mai11}
we adopted a bimodal distribution, with 50\% of the NEOs having
a visible geometric albedo of 0.20 and 50\% having an albedo of 0.055.
This 50:50 ratio holds for NEOs of a given physical size, as used
in our modeling, rather than for NEOs of a given absolute magnitude.

We can express the NEO photon flux per unit wavelength interval, $F$, at the
telescope as follows:
\begin{equation}
F = N_{\lambda}(T_{\sun},\lambda) p_X P(\alpha)
\left ( {R_\sun \over r} \right )^2
\left ( {R \over \Delta} \right )^2
\end{equation}
where $N_{\lambda}$ is the emission from the Sun, $\lambda$ is the wavelength
of light, $p_X$ is the bond albedo of the NEO,
$P(\alpha)$ is the phase function or brightness as a function of
phase angle ($\alpha$),
$R_{\sun}$ is the radius of
the Sun,
$r$ is the heliocentric distance of the NEO, $R$ is the radius of the NEO,
and $\Delta$ is the geocentric distance of the NEO\null.
We used the IAU-standard phase function and phase integral as described in
\citet{bow89} with a typical value of $G=0.2$ applied to all objects.
The blackbody
emissivity in photons per unit wavelength is given by
\begin{equation}
N_{\lambda} = { 2 \pi c \over \lambda^4 [ \exp ( hc/\lambda k T) -1 ] }
\end{equation}
where $T$ is the effective temperature of the Sun, $\lambda$ is the wavelength
of light, $c$ is the speed of light, $h$ is Planck's constant, and $k$ is
Boltzmann's constant.

The sky background in the model was designed for spacecraft observations
and represents the scattered light from zodiacal dust seen from
Earth orbit.  This level is fainter than the sky brightness seen from
the ground.  We scaled our Earth-orbit value up by a factor of 3 to
correspond, on average, to observations with LSST at a zenith angle
of $\sim$45\mydeg.

Based on the expected site conditions for LSST \citep{ive14}, we chose a value
for the seeing (FWHM of point spread function) of 0.65 arcsec for all LSST
observations.

With LSST, the instantaneous FOR is centered on the zenith, with a solid
angle determined by the chosen minimum elevation angle.  During the course
of each night, the instantaneous FOR sweeps up a considerable solid angle.
The sky region near opposition is accessible much of the time, with nearly
peak sensitivity.  As we go farther from opposition, we have a lower duty
cycle for access.  The zenith angle for those observations will be high
and the sensitivity will be reduced.  The time-averaged FOR decreases in
duty cycle (accessibility) and sensitivity as we go away from opposition,
in addition to the sensitivity loss due to sky brightness.  In order to
use the Sentinel performance model, we approximated this tapered region
with a sharp cutoff at $\pm$110\mydeg\ in right ascension from opposition
and from $-$90\mydeg to $+$30\mydeg in declination.

\subsection{Validation of LSST model predictions}

With these representations of NEA flux and sky brightness in visible
wavelengths, we calculated the performance of the ground-based
visible-light LSST telescope for NEA detections.  LSST has presented
analyses that show it is expected to reach 84\% completeness in 10 years
on 140-meter PHAs with its nominal observing cadence \citep{ive07}.
Completeness levels will be slightly lower for the NEA population,
similar to what is seen with the Sentinel analysis.  Their measure of
completion refers to objects with at least two observations in at least
three nights within one lunation.  For our Sentinel models we require
that the observations extend over more than a Sentinel observing cycle
of 30 days in order to provide a sufficiently high quality orbit for
future follow-up observations.  We apply that same cataloging criterion
to the LSST modeling. In our modeling of LSST, a detection consists of
4 $r$-band observations in a lunation and we require detections in two
separate lunations for an object to be considered cataloged.  The time
it takes to cover the LSST FOR four times in a lunation is less than the
total time available.  This time margin allows for observations with
other filters or cadences of lower value for NEA searches.  Our model
is not sensitive to the exact temporal pattern.  These detection and
cataloging conditions are different from those used by \citep{ive07}.
However, we feel our choices are reasonable approximations that are
easily implemented with our tools.

Our model results for LSST performance is a 73\% catalog completeness
for NEAs larger than $D=$140~m in 10 years of observations.  This is
substantially lower than the 82\% obtained by the LSST team (www.lsst.org).
The difference is primarily due to the differing albedo assumptions.
We assumed 50\% of NEAs of a given diameter have 5.5\% albedo and 50\%
have 20\% albedo as suggested by \citep{bot02}.  The LSST team used a
uniform 14\% albedo.  When we use the LSST albedo assumption,
we predict an 80\% catalog completeness.

\subsection{Sentinel performance in combination with ground surveys}

Both Sentinel and LSST are shown to be highly capable search programs for
NEAs.  There will be many discoveries common to both the space and ground
search programs, and it is important to examine this redundancy in surveys
to assess the value of the addition of a space observatory.  Redundancy is
valuable as both confirmation and extra assurance that objects will not
be overlooked but will degrade the combined NEA discovery rate.  At the
smallest sizes, Sentinel’s discoveries will be largely uncorrelated
with Earth-based detections because, in its heliocentric orbit,
Sentinel will observe a different volume of space with unique new NEAs.

In order to compare the incremental advantage of an additional survey,
it is necessary to move beyond mere completion statistics to a capability
that includes recognizing each discovery as an individual and removing
redundant observations of known objects from the second survey. We have
performed this analysis for Sentinel using our model of LSST performance
and assuming concurrent operation of Sentinel and LSST (2022 start).

Results of our simulations are given in Table~\ref{tbl-dual}.  The
Sentinel mission has a nominal duration of 6.5 years and the LSST project
has a nominal survey duration of 10 years.  The survey simulations are
linked so that we know which objects are common or unique to both surveys.
Cumulative completion values are provided for NEAs and virtual impactors for
three different limiting sizes.  The last three columns give the fraction
of the total return that comes from Sentinel alone ($f_{\rm S}$), LSST
alone ($f_{\rm L}$), and seen by both ($f_{\rm S+L}$).  As seen in this
table, the correlation between the samples of the two surveys is
highest for $D>140$~m objects and decreases as the limiting size decreases.
At the smallest sizes, Sentinel clearly provides a significant increase
in the number of objects cataloged that are not seen by LSST, regardless of
what category of object is considered.

\begin{deluxetable}{ccccccc}
\tablecaption{Completeness for Sentinel$+$LSST\label{tbl-dual}}
\tablewidth{0pt}
\tablehead{
\colhead{Population}&
\colhead{S(6.5)}&
\colhead{L(10)}&
\colhead{S$+$L}&
\colhead{$f_{\rm S}$}&
\colhead{$f_{\rm L}$}&
\colhead{$f_{\rm S+L}$}
}
\startdata
NEA  (D$>$140m)& 82\%& 69\%& 91\%& 24\%& 10\%& 66\%\cr
NEA   (D$>$40m)& 27\%& 15\%& 34\%& 56\%& 21\%& 23\%\cr
NEA   (D$>$20m)&  8\%&  4\%& 10\%& 60\%& 20\%& 20\%\cr
VIMP (D$>$140m)& 93\%& 83\%& 97\%& 14\%&  4\%& 82\%\cr
VIMP  (D$>$40m)& 57\%& 35\%& 67\%& 48\%& 15\%& 37\%\cr
VIMP  (D$>$20m)& 23\%& 13\%& 30\%& 57\%& 23\%& 20\%\cr
\enddata
\tablecomments{\scriptsize
This table contains the survey performance for Sentinel alone for 6.5 years,
S(6.5); LSST along for 10 years, L(10); and both surveys for their respective
durations, S+L.  The total number detected in each case is further broken
down into the amount seen by Sentinel only ($f_{\rm S}$), seen only by LSST
($f_{\rm L}$), and seen by both facilities ($f_{\rm S+L}$).
}
\end{deluxetable}

% NS  = ns + nsl
% NL  = nl + nsl
% NSL = ns + nl + nsl
%
% ns  = NSL - NL
% nl  = NSL - NS
% nsl = NSL - ns - nl
%
% fs  = ns/NSL
% fl  = nl/NSL
% fsl = nsl/NSL

% This is rolled up into fracs.pro

\begin{center}
\includegraphics[scale=0.29,trim=80 30 50 30]{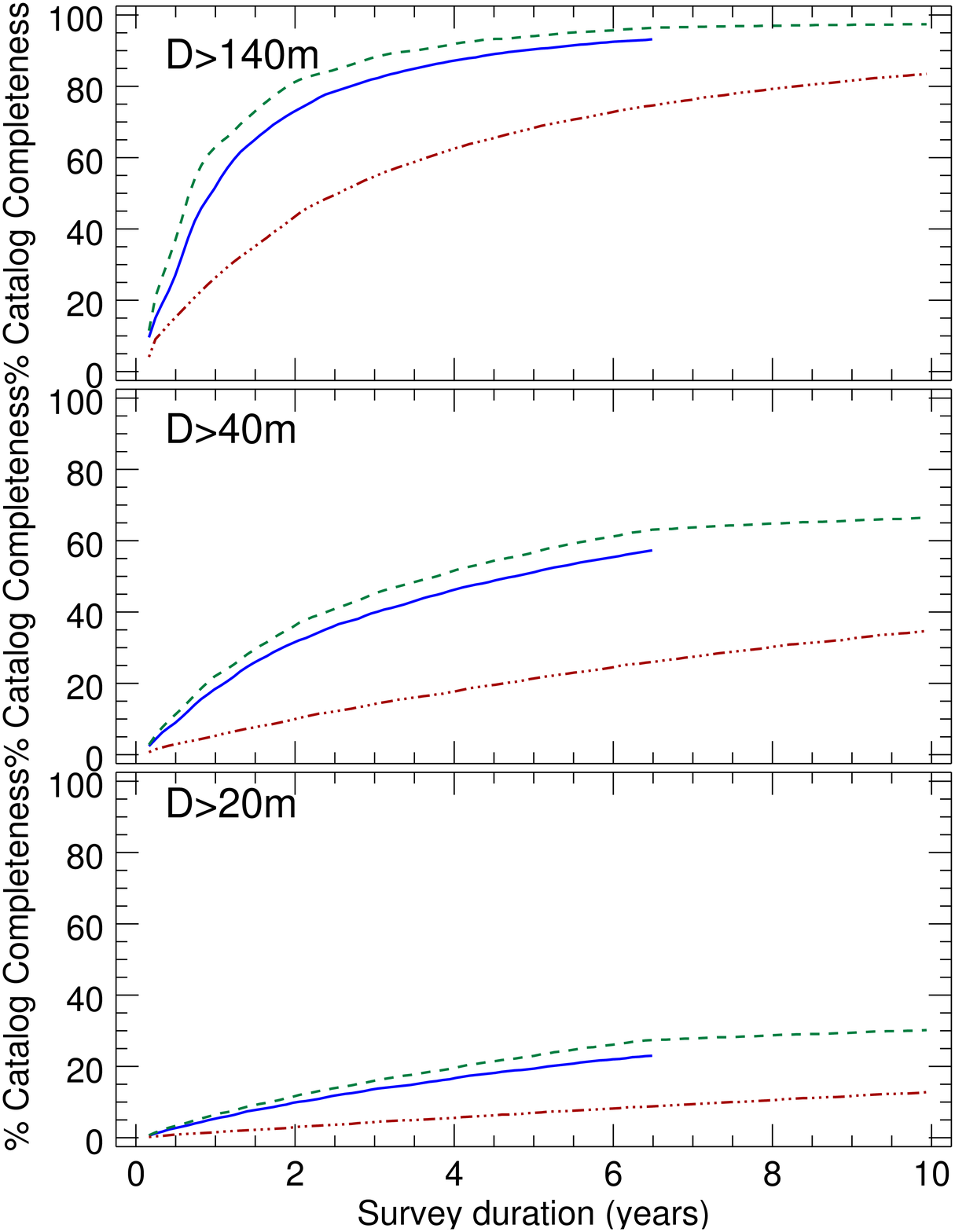}
\figcaption[figlsstcomp.eps]{\label{fig-lsstcomp}\scriptsize
Survey completion rates for virtual impactors at different limiting
sizes.  Shown are completion curves for the baseline Sentinel mission
(blue solid lines), LSST (dot-dash red line), and Sentinel and LSST
together (green dashed lines).  The Sentinel survey only runs for 6.5
years while LSST is shown for 10 years.
}
\end{center}

Figure~\ref{fig-lsstcomp} provides a graphical summary of the joint
performance for these three sizes.  Sentinel, designed as a dedicated NEA
survey system, has a higher performance level than LSST.  Sentinel added
to LSST significantly increases the completeness achieved compared
to LSST alone.  For NEAs larger than 140 meters Sentinel gets 84\%
completion and LSST also does quite well with 73\% completion but the
combined performance reaches 93\%.  As the target size decreases the
gap widens somewhat.  The redundancy nearly vanishes between these two
facilities as the target size shrinks.  For the impactor population
the performance is even better with the combined survey reaching 98\%
completeness for $D>140$~m.  For the smallest size considered here
of $D>20$~m, the combined survey can reach a completeness of 34\% for
cataloging impactors and more than half of those are found by Sentinel
alone.

\begin{deluxetable}{ccccccc}
\tablecaption{Sentinel, 1~AU, FOR=66\mydeg -140\mydeg\ with and without LSST\label{tbl-for65}}
\tablewidth{0pt}
\tablehead{
\colhead{Population}&
\colhead{S$'$(6.5)}&
\colhead{L(10)}&
\colhead{S$'$+L}&
\colhead{$f_{{\rm S}'}$}&
\colhead{$f_{\rm L}$}&
\colhead{$f_{{\rm S}'+{\rm L}}$}
}
\startdata
Imp D$>$140m& 86\%& 83\%& 92\%& 10\%&  7\%& 83\%\cr
Imp  D$>$40m& 45\%& 35\%& 54\%& 35\%& 17\%& 48\%\cr
Imp  D$>$20m& 18\%& 13\%& 24\%& 46\%& 25\%& 29\%\cr
\enddata
\tablecomments{\scriptsize
This table contains survey performance for a modified Sentinel operating
at 1 AU looking mostly at quadrature combined with LSST\null.  The cumulative
completeness is shown
Sentinel alone for 6.5 years,
S$'$(6.5); LSST along for 10 years, L(10); and both surveys for their respective
durations, S$'$+L.
The total number detected in each case is further broken
down into the amount seen by Sentinel only ($f_{{\rm S}'}$), seen only by LSST
($f_{\rm L}$), and seen by both facilities ($f_{{\rm S}'+{\rm L}}$).
}
\end{deluxetable}

We used this model to also assess the redundancy of detections by an
IR observatory in the Earth-Sun L1 point with those by ground-based
surveys.  This case is basically the ``N'' or NEOCam option of Sentinel
with a modified FOR that covers 66\mydeg\ to 140\mydeg solar elongation.
Table \ref{tbl-for65} shows the result of this calculation for impactors.
The combined performance is slightly worse than for the baseline Sentinel
case with the combined survey reaching 93\% for $D>140$~m objects.
Extending down to 20~m objects, the combined completeness drops to
27\% down from 34\% for the baseline Sentinel case.  The difference is
largely due to the redundancy of space discoveries with the discoveries
already made by LSST when the space observatory is located near the Earth.

\section{Discussion}\label{sect-disc}

These tools and results are very useful for establishing survey
performance over a range of operating conditions.  An important outcome
is the ability to provide quantitative performance metrics for different
design options.  A very strong guiding principle for the survey is to
generate a catalog of objects with high-quality orbits and maximize being
able to link other detections with other data.  This is a very different
strategy from a survey that is optimized to obtain the first detection.
Such a narrowly focused first-detection survey requires additional
resources to handle the large volume of followup work required for
quality orbit determination.  A survey that is designed to do its own
followup, such as Sentinel, need not be optimized by the same design
strategies as a first-detection survey.  Indeed, our results indicate
that optimizing the field-of-regard to enhance first detection (guided
by Figs.~\ref{fig-neo-v}-\ref{fig-vimp-e}) does not yield a better survey
outcome when measured against multi-epoch observations.  Such a difference
is a consequence of the need for a self-followup survey to find the
object at somewhere other than the place where it is most easily seen.
Our attempts to optimize the survey based on the FOR show that even the
low-density regions near the ecliptic poles have an important contribution
to the survey output.  The time that might be saved for not observing in
those locations only serves to decrease the time between observations
that you would already get.  The only optimization that is important
is in adapting for the preference for opposition or quadrature which is
also coupled to the orbit chosen for the observatory.

Considerable effort was given to understanding our small-object
performance and looking for ways to improve the survey in this area.
Along the way it was very interesting to see that the large object
($D>140$~m) performance was nearly as good with all the options we tried.
Variations in strategy that included changing cadence, exposure time,
telescope aperture, FOR, and FOV, all gave essentially the same answer:
good coverage but not quite the 90\% coverage desired during the nominal
mission duration.  The most effective thing for better completeness for
large objects is to run the survey for longer.  A far more important
consideration comes from defining the goals of the survey.  The hardest
population to get 90\% completion on is the NEA group due to their
diversity of orbital properties.  It is this result that leads to
differing goals for this type of survey.  For example, here are a few
distinct goals: 1) deduce the population of asteroids in the near-Earth
region of the solar system, 2) find all of the asteroids in the near-Earth
region, or 3) find all of the objects that pose an imminent threat to
the Earth.  This is by no means an exhaustive set of possible goals but
serves to frame the discussion.

Case \#1, elucidating the population, is perhaps the easiest task of
the three since a survey only needs to find enough objects such that
the error introduced during the de-biasing of the survey data are small
enough for the subsequent scientific investigation.  Arguably, we already
have enough objects cataloged to permit successful de-biasing of the
survey to directly constraint our knowledge of the
population in its overall properties.  Uncertainties of a factor of a
few may still be possible down to $D>20$~m and there are still questions
about the actual shape of the distribution but the Earth impact record
gives us very important ground-truth for the modern era.  Also, this
case has no particular urgency for its completion.  Given enough time,
efforts from relatively inexpensive ground-based efforts will eventually
solve this particular problem.  Indeed, most of what we know now comes
from ground-based efforts.

Case \#2, finding NEAs down to a limiting size, is a difficult task
that is limited nearly as much by orbital properties as the
detection strategy.  The full NEAs orbit distribution includes objects
with either long periods or other complicating geometric constraints
that leave them impossible to detect almost all of the time without
extreme effort.  In this case, reaching a 90\% completeness level will
require survey durations matched to the orbital outliers.  Furthermore,
the vast majority of these objects bear no threat against the Earth.
Again, if the goal is a scientific investigation of the population,
completeness is not required.  Completeness is a compelling issue when
weighed against the threat of impact on the Earth.  This argument brings
us to case \#3.

If the goal of NEA surveys is to identify actual threats to the Earth,
such surveys should be optimized for this outcome.  Clearly, with such
a focus, an object that won't hit the Earth in the next billion years
should be of no concern when designing the survey.  However, the George
E\null. Brown act survey mandate does just that by ``requiring'' detection
of NEAs for the mandate to be fulfilled, all in the name of identifying
threats to the Earth.

A survey dedicated to threatening objects takes on a very different
form from a more general NEA survey.  The NEA population can be
split up into three distinct groups.  The first group contains those
objects that {\em will} hit the Earth in the next 100 years -- call
these objects the set of impactors.  The second group contains objects
that will become an impactor 100 years or more in the future up to some
limiting timescale of interest -- call these the set of future impactors.
The third group is comprised of those objects posing no threat at all
-- call these the background population.  The choice of 100 years for
the time horizon of concern is chosen to loosely match the timescale
for chaotic evolution of the NEA orbits.  Within a 100 year period
from now we can, with sufficient data, accurately predict an object's
motion and know with certainty that it either will or will not impact
the Earth.  Beyond 100 years the chaotic nature of the orbits precludes
firm knowledge, only a likelihood of impact.  This 100 year window is
perhaps not rigorously correct but serves as a commonly used guideline
for the prediction threshold.  For the purposes of this discussion
we consider the 100 year limit to be a hard cutoff even though the real
limit depends on the actual object and its size, not to mention the
quality of its orbit estimate.

The set of impactors is very small and is a reasonably well known
number just from recent history.  As shown in Fig.~\ref{fig-impact},
current estimates indicate there are 7 objects down to 15~m that will
impact the Earth in the 100 years following the survey.  The set of
future impactors would thus be 700 if the time-horizon of concern were
set to be 100 centuries.  That is still a small number.   Of course,
this set cannot be deterministically known.  The chaotic evolution of
the impactor orbits will dictate a mean probability of impact for this
group and those that will be close to, but not actual impactors.  If,
for example, the mean probability were as low as 10\% in this group,
then the sample of objects worth tracking would then be 7000 objects.
In reality, this probability will evolve from being deterministic at 100
years out to being ever less certain as you look further into the future.
These two sets of objects are in stark contrast by numbers against
the third category of background NEAs.  According to recent estimates
by \citet{bos15}, there are nearly 100 million objects with $D>15$~m.
The overwhelmingly vast majority of objects out there to be found are
those that will not ever strike the Earth.  Our analysis shows there are
ways to bias a survey toward improved efficiency of finding impactors
at the expense of not finding as many other non-threatening types.

One result that has become very clear during this analysis is just how
big the problem is for finding impactors.  There are many steps along the
way toward planetary defense.  The first step is a wide-field survey.
Ideally, surveys should continue until all the objects are found but
subsequent steps can begin before the survey process is completed.
The output of the survey process is a catalog of objects with orbits of
varying quality.   These orbits will range from very well determined
to very poorly determined.  For any objects with a non-zero impact
probability, we need to continue with long-term followup observations.
Additional observations will reduce the list of candidate impactors.
This followup work is distinguished from the initial survey by the need
to re-observe an already known object.  Some objects will be well enough
known to permit a targeted observation while others will effectively
require rediscovery.  In the end, the requirement is to get a minimum
astrometric arc that will prune the candidate list down from 10$^8$
to something manageable for concerted high-precision work.  The details
of how long the observational arc must be are beyond the scope of this
work but should be addressed, especially in light of what the astrometric
data quality will be like once the Gaia star catalog becomes available.
It may well be that completing this initial cataloging step may require
two separate all-sky surveys separated by some period of time, say 10
years, to ensure long-arc orbital constraints on everything.

The final step comes when one of the candidate list is then found likely
to be on a collision course with Earth.  From there the mitigation work
begins with a definite goal.  Given the nature of the size distribution
of NEAs, very small objects will likely be the first identified.
The discovery of a small Chelyabinsk-sized object would actually be the
ideal for our first attempt at space-based mitigation since the impact
effects can be protected against by relatively simple civil defense
actions local to the affected area.  The inescapable conclusion here
is that planetary defense is a long-term task and really has no end.
The startup phase of getting complete catalogs is the hardest part and
will require many decades of effort, or longer.  The Sentinel mission we
have studied here is a big step along the way.  By combining its efforts
with a complementary LSST effort we can improve the cataloging rate and
reach a sample size that is likely to net the first actual impactor.

\section{Conclusions}

We have looked at many variations for Sentinel and find that, for small
impactors, the nominal orbit provides a very effective location for
surveying and we do not find any options for substantial improvement
that do not markedly increase the mission cost -- aperture of the
telescope being the most important.  In particular, Sentinel can reach
50\% completeness for impactors larger than 40 meters.  Observatory
location is not critical for any sub-population except for impactors
and ARRM targets. In particular, observing from an Earth-neighborhood
location is inferior to the interior orbit for objects of interest to
planetary defense or human exploration targets.  One clear lesson from
this work is the need to be concerned with modeling performance down to
a $D=15-20$~m  size range for these two goals.  For planetary defense,
this conclusion is supported by the fact that most near-term threats
come from the smallest objects and you have to get down to this small
size before an event is likely in the next 100 years.

Depending on the current completeness level of any population,
optimization of a survey could lead to changes in the survey design if
only the undiscovered population were considered.  If a survey were merely
an incremental improvement, this could be a very important consideration.
Our survey modeling leads to guidance on an optimized strategy assuming
you haven't yet found any objects.  The real situation is somewhere
in the middle between these two cases for Sentinel but closer to the
``new'' survey end of the spectrum due to working to much smaller sizes
and working in the infrared where the biases against finding low albedo
objects are removed.  Modeling the undiscovered sub-populations will be
left to future work, if warranted.

Working to such small sizes leads to an fundamentally higher discovery
rate than is currently taking place.  The reality of higher discovery
rates requires significant effort for followup as well and new surveys
really must work to do their own followup as much as possible.  In our
survey design we also make a strong distinction between detection and
cataloging.  When modeling survey performance it is also important
to track the observational arc that would be obtained on any newly
discovered object.  There is a minimum arc required to ensure linking
against other later observations and this point is often overlooked.
For our survey performance metrics we insist on a minimum arc of 28 days
but this value is more of a guideline than a hard requirement.  The
minimum required arc needed to claim an object is cataloged might be
worth additional study.

Object detection is indeed easier in some locations in the sky, --
the so-called ``sweet spots'' where the probability of detection
is at its peak \citep[cf.,][]{che04,ver09}.  As shown in Figures
\ref{fig-neo-v}--\ref{fig-vimp-e}, the discovery sweet spot depends on
the observatory location and it also depends on object size and aperture.
It is clear from our analysis that the best detections of an object will
most likely come from these areas of the sky but the minimum arc length
goal pushes the followup observations into other regions of the sky.
For this reason, the FOR of a survey that does its own followup must
necessarily cover more sky than just the sweet spots.

NEA surveying is a difficult enough task, particularly at the small size
range, that cost effective survey plans need to consider combining efforts
from other facilities.  This combined survey strategy can be accomplished
in many ways.  One easy example is to fly more than one space observatory
and placed in different locations in space.  For instance, a set of
three Sentinel telescopes in a Venus-like orbit spaced by 120\mydeg\
of longitude would be three times as fast at finding the objects down
to $D=15$~m.  This group would also be much more effective at followup
and give much longer observational arcs for the resulting object orbit
catalog.  Such strategies are very powerful but also quite expensive.
In the near term, any such space-based facility really should work to
be complementary to ground-based surveys, represented in our study by
LSST so that both surveys can add the largest number of objects to our
NEA catalogs.  In the case of Sentinel plus LSST, we find the combined survey
will find more than 70\% of asteroid impactors larger than 40 meters.
The choice of orbit for a survey telescope also an important consideration.

Our modeling of the combined Sentinel plus LSST performance will be a
catalog of perhaps 1000's of possible impactors that will require
continued, potentially long-term followup observations to sift out
the real impactors from close misses.  Additionally, this catalog will
permit routine prediction of future close passes for continued study,
significantly reducing the number of objects that will sneak up on the
Earth without warning before its first detection.

We are poised on the threshold of a huge step forward in NEA surveys,
a step that will begin to reveal a more complete picture of the
objects in space around us.  This step will come from a space-based
approach combined with ground-based observations, especially LSST\null.
The task of mitigating against impacts does not end with this step.
Continued surveys will be required for a while to get good observational
arcs on all discovered objects as well as continuing to strive for ever
better completeness for small objects.  This initial phase of catalog
will likely require at least two such survey efforts separated in time.
After that work, followup becomes ever more increasingly targeted and
no longer requires all-sky surveys.  Nonetheless, protecting the Earth
against impactors is a long-term and unending task and doing so is
clearly within our grasp.

\acknowledgments
{\footnotesize
Funding for this work provided by the B612 Foundation and its Founding
Circle donors (K. Algeri-Wong, B. Anders, G. Baehr, W. K. Bowes
Jr\null. Foundation, B. Burton, A. Carlson, D. Carlson, S. Cerf, V. Cerf,
Y. Chapman, A. Denton, E. Dyson, A. Eustace, S. Galitsky, E. Gillum,
L. Girand, Glaser Progress Foundation, D. Glasgow, J. Grimm, S. Grimm,
G. Gruener, V. K. Hsu \& Sons Foundation Ltd., J. D. Jameson, J. Jameson,
M. Jonsson Family Foundation, S. Jurvetson, D. Kaiser, S. Krausz,
J. Leszczenski, D. Liddle, S. Mak, G. McAdoo, S. McGregor, J. Mercer,
M. Mullenweg, D. Murphy, P. Norvig, S. Pishevar, R. Quindlen, N. Ramsey,
P. Rawls Family Fund, R. Rothrock, E. Sahakian, R. Schweickart, A. Slater,
T. Trueman, F.B. Vaughn, R.C. Vaughn, B. Wheeler, Y. Wong, M. Wyndowe plus
8 anonymous donors).  Thanks to Lowell Observatory and E. L. G. Bowell
and L. H. Wasserman for the use of the astorb.dat database, funded by
NASA grant NAGW-1470 and the Lowell Observatory endowment.
}

\clearpage


\begin{thebibliography}{}

\bibitem[Bottke {\it et al.}(2002)]{bot02}
Bottke, W.F., Morbidelli, A., Jedicke, R.,  Petit, J.-M., Levison, H.,
Michel, P., and Metcalfe, T.S. 2002.
Icarus 156, 399.

\bibitem[Boslough {\it et al.}(2015)]{bos15}
Boslough, M., Brown, P., and Harris, A. 2015.
IEEE Aerospace Conference, Big Sky, MT.
doi 10.1109/AERO.2015.7119288

\bibitem[Bowell \etal(1989)]{bow89}
Bowell, E., Hapke., B., Domingue, D., Lumme, K., Peltoniemi, J.
and Harris., A. W. 1989.
Asteroids II, eds. R.P. Binzel, T. Gehrels, M.S. Matthews, Univ. of Arizona
Press, pp 524-556.

\bibitem[Brown \etal(2013)]{bro13}
Brown, P.G., Assink, J.D., Astiz, L.,  et al. 2013.
Nature 503, 238.

\bibitem[Cellino \etal(2004)]{cel04}
Cellino, A., Muinonen, K., \& Tedesco, E. F. 2004,
Advances in Space Research, 33, 1576.

\bibitem[Chesley \& Spahr(2004)]{che04}
Chesley, S.R., Spahr, T.B., 2004.
Earth-impactors: Orbital characteristics and warning times.
In: Belton, M.J.S., Morgan, T.H., Samarashinha, N.H., Yeomans, D.K. (Eds.),
Mitigation of Hazardous Comets and Asteroids. Cambridge University Press,
Cambridge, pp. 22-37.

%\bibitem[Denneau {\it et al.}(2013)]{den13}
%Denneau, L., Jedicke, R., Grav, T., et al. 2013, PASP, 125, 357.

%\bibitem[Gehrels(1986)]{geh86}
%Gehrels, T., 1986. In: Lagerkvist, C.-I., Lindblad, B.A., Lundstedt, H.,
%Rickman, H. (Eds.), Asteroids, Comets, and Meteors, vol. II.
%University of Arizona Press, Tucson, pp. 19.

\bibitem[Granvik {\it et al.}(2009)]{gra09}
Granvik, M., Virtanen, J., Muinonen, K. 2009.
M\&PS 44, 1853.

%\bibitem[Harris \& D'Abramo(2015)]{har15}
%Harris, A. W. \& D'Abramo, G. 2013.
%Icarus 257, 302.

\bibitem[Ivezi\'c {\it et al.}(2007)]{ive07}
Ivezi\'c, Z., J. A. Tyson, M. Juri\'c, J. Kubica, A. Connolly,
F. Pierfederici, A. W. Harris, E. Bowell, and the LSST
Collaboration. 2007. LSST: Comprehensive NEO Detection, Characterization,
and Orbits. In Near Earth Objects, Our Celestial Neighbors: Opportunity
and Risk Proceedings International Astronomical Union Proceedings,
edited by A. Milani, G. B. Valsecchi and D. Vokrouhlick\'y.

\bibitem[Ivezi\'c {\it et al.}(2006)]{ive06}
Ivezi\'c, Z., Tyson, J. A., Juri\'c et al. 2006.
IAU Symposium 236 ``Near Earth Objects, Our Celestial Neighbors: Opportunity
and Risk'', Proc IAU Symp No. 236, p. 353.
%http://astro-ph/0701506.pdf

\bibitem[Ivezi\'c {\it et al.}(2014)]{ive14}
Ivezi\'c, Z., Tyson, J. A., Abel, B. et al. 2014.
http://arxiv.org/pdf/0805.2366.pdf

\bibitem[Jedicke(1996)]{jed96}
Jedicke, R. 1996
AJ 111, 970.

\bibitem[Jedicke \etal(2003)]{jed03}
Jedicke, R., Morbidelli, A., Petit, J., 2003.
Icarus 161, 17–33.

\bibitem[Jedicke \etal(2006)]{jed06}
Jedicke, R., Magnier, E.A., Kaiser, N., Chambers, K.C., 2006.
Proceedings IAU Symposium No. 236. Cambridge University Press,
Cambridge, Milani, A., Valsecchi, G.B., Vokrouhlicky´ ,
D. (Eds.), pp.341–352.

\bibitem[Jedicke \etal(2015)]{jed15}
Jedicke, R., Granvik, M., Micheli, M., Ryan, E., Spahr, T.,
and Yeomans, D. K., 2015.  In {\it Asteroid IV} (P. Michel \etal, eds.)
pp. 795-813.

\bibitem[Kelsall {\it et al.}(1998)]{kel98}
Kelsall, T., Weiland, J. L., Franz, B. A., Reach, W. T., Arendt, R. G.,
Dwek, E., Freudenreich, H. T., Hauser, M. G., Moseley, S. H., Odegard,
N. P., Silverberg, R. F., and  Wright, E. L. 1998.
Apj 508, 44.

\bibitem[Koch \etal(2010)]{koc10}
Koch, D.G., Borucki, W.J., Basri, G., Batalha, N.M., Brown, T.M.,
Caldwell, D., Christensen-Dalsgaard, J., Cochran, W.D., DeVore, E.,
Dunham, E.W., Gautier, T.N., III, Geary, J.C., Gilliland, R.L., Gould,
A., Jenkins, J., Kondo, Y., Latham, D.W., Lissauer, J.J., Marcy, G.,
Monet, D., Sasselov, D., Boss, A., Brownlee, D., Caldwell, J., Dupree,
A.K., Howell, S.B., Kjeldsen, H., Meibom, S., Morrison, D., Owen,
T., Reitsema, H., Tarter, J., Bryson, S.T., Dotson, J.L., Gazis,P.,
Haas, M.R., Kolodziejczak, J., Rowe, J.F., Van Cleve, J.E., Allen, C.,
Chandrasekaran, H., Clarke, B.D., Li, J., Quintana, E.V., Tenenbaum,
P., Twicken, J.D., Wu, H. (2010).  ApJL 713, L79.

\bibitem[Larson(2007)]{lar07}
Larson, S. 2007, in IAU Symp. 236, Near Earth Objects, our Celestial
Neighbors: Opportunity and Risk, ed. G. B. Valsecchi, \& D. Vokrouhlick\'y,
(Cambridge: Cambridge University Press), 323.

\bibitem[Lu {\it et al.}(2013)]{lu13}
Lu , E.T., Reitsema, H.J. Troeltzsch, J. Hubbard, S. 2013.
New Space 1, 425.

\bibitem[Mainzer {\it et al.}(2015a)]{mai15a}
Mainzer, A., Grav, T., Bauer, J. {\it et al.} 2015a.
AJ 149, 172.

\bibitem[Mainzer {\it et al.}(2015b)]{mai15b}
Mainzer, A., Bauer, J., Grav, T., et al. 2015b.
Space-Based Infrared Discovery and Characterization
of Minor Planets with NEOWISE. In Handbook of Cosmic Hazards and Planetary
Defense, Edited by  J. Pelton and F. Allahdadi,
Springer Intl. Pub, Switzerland,
p. 583-611.

\bibitem[Mainzer \etal(2011)]{mai11}
Mainzer, A., Grav, T., Bauer, J. et al. 2011.
ApJ 743 156.

\bibitem[NRC Committee(2010)]{nrc10}
NRC Committee to Review Near-Earth Object Surveys and Hazard Mitigation
Strategies; National Research Council, 2010.  Defending planet Earth:
near-Earth object surveys and hazard mitigation strategies, 152 pp.
ISBN 978-0-309-14968-6

\bibitem[Pravdo \etal(1999)]{pra99}
Pravdo, S.H., Rabinowitz, D.L., Helin, El.F., and 11 co-authors, 1999.
AJ 117, 1616.

\bibitem[Pravec {\it et al.}(2012)]{pra12}
Pravec, P., Harris, A.W., Ku{\v s}nir\'ak, P., Gal\'ad, A.,
and Hornoch, K. 2012.
Icarus 221, 365.

%\bibitem[Reitsema \& Buie(2015)]{rei15}
%Reitsema, H.J. \& Buie, M.W. 2015.
%4th IAAA Planetary Defense Conference, Frascati, Italy.

\bibitem[Reitsema \etal(2014)]{rei15}
Reitsema, H.J. and Lu, E.T. 2015.
Sentinel: A Space Telescope Program to Create a 100-Year Asteroid
Impact Warning. In Handbook of Cosmic Hazards and Planetary Defense,
Edited by J. Pelton and F. Allahdadi, Springer Intl. Pub, Switzerland,
pp. 570-581.

\bibitem[Spencer(1989)]{spe89}
Spencer, J.R., Lebofsky, L.A., \& Skyes, M.V. 1989.
Icarus 78, 337.

\bibitem[Spencer(1990)]{spe90}
Spencer, J.R. 1990.
Icarus 83, 27.

\bibitem[Stokes \etal(2000)]{sto00}
Stokes, G.H., Evans, J.B., Viggh, H.E.M., Shelly, F.C., Pearce, E.C., 2000.
Icarus 148, 21.

\bibitem[Tedesco \etal(2000)]{ted00}
Tedesco, E.F., Muinonen K., and Price S.D. (2000).
Planet. Space Sci. 48, 801.

\bibitem[Vere{\v s} {\it et al.}(2009)]{ver09}
Vere{\v s}, P., Jedicke, R., Wainscoat, R., et al.\ 2009, Icarus, 203, 472.

\bibitem[Willson {\it et al.}(1980)]{wil80}
Willson R.C., C.H. Duncan, and J. Geist 1980,
Science 207, 177.

\end{thebibliography}
\end{document}